\documentclass[aps,prd,twocolumn,superscriptaddress,nofootinbib,10pt]{revtex4-2}

\usepackage[T1]{fontenc}
\usepackage{amsmath,amssymb,bm,mathtools}
\usepackage{hyperref}
\usepackage{xcolor}
\usepackage{microtype}
\usepackage{url}
\usepackage{orcidlink}
\definecolor{linknavy}{RGB}{25,55,95}
\definecolor{linkred}{RGB}{150,40,55}
\hypersetup{
    colorlinks         = true,
    linkcolor          = blue,
    citecolor          = red,
    urlcolor           = linknavy,
    filecolor          = linknavy,
    menucolor          = linknavy,
    pdfstartview       = FitH,
    pdfpagemode        = UseOutlines,
    bookmarks          = true,
    bookmarksnumbered  = true,
    bookmarksopen      = true,
    bookmarksopenlevel = 2,
    pdfdisplaydoctitle = true,
    pdftitle            = {Optical Geometry of a Lorentz-Violating BTZ-Like Black Hole: Reflection Double and Dirac Scattering},
    pdfauthor           = {Abdullah Guvendi and Omar Mustafa},
    breaklinks         = true
}
\allowdisplaybreaks[1]
\newcommand{\dd}{\mathrm{d}}
\newcommand{\sech}{\operatorname{sech}}

\newcommand{\Lopt}{\ell_{\mathrm{opt}}}

\begin{document}

\title{Optical Geometry of a Lorentz-Violating BTZ-Like Black Hole: Reflection Double and Dirac Scattering}

\author{Abdullah Guvendi\orcidlink{0000-0003-0564-9899}}
\email{abdullah.guvendi@erzurum.edu.tr}
\affiliation{Department of Basic Sciences, Erzurum Technical University, 25050 Erzurum, T\"urkiye}

\author{Omar Mustafa\orcidlink{0000-0001-6664-3859}}
\email{omar.mustafa@emu.edu.tr}
\affiliation{Department of Physics, Eastern Mediterranean University, 99628, G. Magusa, North Cyprus, Mersin 10, T\"urkiye}

\date{November 4, 2026}

\begin{abstract}
We analyze the optical geometry of a static BTZ-like black-hole metric with a Lorentz-violating radial factor, distinguishing the one-sided black-hole exterior from the reflection double constructed from its optical spatial section. The optical metric is ultrastatic and has constant negative Gaussian curvature. The reflection double is smooth and complete, with a minimal circle; it is a geometric completion, not a demonstrated solution of the Einstein--bumblebee equations. We derive exact criteria for null-geodesic crossing and return, evaluate the corresponding angular sweeps, and determine the finite domain of a Euclidean surface-of-revolution embedding. Two inequivalent magnetic configurations are specified as external test fields on the embeddable patch. For a continuum graphene analogue on the doubled geometry, we derive the covariant massless Dirac operator and reduce its radial equation to Scarf-II partner equations. The full-line connection problem yields flux-normalized complex scattering amplitudes and probabilities, the resonance-pole lattice and its gamma-function residues, spectral widths and quality factors, the digamma Wigner delay, and conditional pole contributions to packet dynamics. We compare the analytic amplitudes with direct numerical integration of the radial partner equation, including their complex phases. We further separate fixed-$\ell$ from fixed-$\Lambda_e$ parameter protocols and audit the optical effective-mass degeneracy using Lorentzian curvature invariants, timelike geodesics, and a massive Klein--Gordon probe. Throughout, the gravitational completion, prescribed fields, and continuum scattering results are interpreted only within the scope of their respective constructions. Device-level graphene transport and a self-consistent gravitational wormhole require additional physical input.
\end{abstract}


\maketitle

\section{Introduction}\label{sec:intro}
The Bañados--Teitelboim--Zanelli (BTZ) black hole is a fundamental exact solution of $(2+1)$-dimensional gravity with negative cosmological constant \cite{BTZ,BTZ2}. Its relatively simple geometry has made it a useful setting for studying black-hole thermodynamics, horizon structure, and anti-de Sitter gravity \cite{BTZ,BTZ2,BTZEntropy,Birmingham}. The importance of the BTZ solution in three-dimensional gravity is also reflected in the asymptotic-symmetry framework of Brown and Henneaux and in reviews of lower-dimensional gravity and black holes \cite{BrownHenneaux,CarlipReview,Mann}. Beyond gravitational theory, BTZ-related geometries have also been used in effective descriptions of wave propagation and condensed-matter systems.

Optical geometry provides a direct link between static gravitational backgrounds and spatial ray dynamics. Conformal rescaling by the inverse lapse maps spatial projections of null geodesics to geodesics of a Riemannian optical metric \cite{Wald}. The geometric and embedding properties of optical metrics have been developed for static and stationary spacetimes, with the latter admitting a Zermelo--Randers description \cite{GibbonsWarnick,GibbonsHerdeiro}. For BTZ geometry, Cvetič and Gibbons established an optical-geometric connection with embedding methods and graphene-inspired models \cite{CveticGibbons}. Related optical-wormhole constructions have considered ray propagation and wave-optical behavior on extended optical geometries \cite{BTZWormhole,OpticalWormhole}. These studies provide the setting for examining optical structures generated by deformations of the underlying spacetime metric.

Lorentz-symmetry breaking in gravity is commonly formulated through vector--tensor theories in which a vector field acquires a nonzero vacuum expectation value \cite{KosteleckySamuel,KosteleckyPotting}. Reviews describe the theoretical framework and phenomenology of Lorentz violation, while compilations summarize experimental and observational bounds \cite{Bluhm,KosteleckyRMP}. Gravitational effects of Lorentz-violating couplings have been investigated in post-Newtonian and vacuum settings \cite{BaileyKostelecky,BertolamiParamos}; black-hole and related solutions include Schwarzschild-like geometries, cosmological extensions, and perturbative analyses \cite{Casana,MalufNeves,MaiXuLiangShao,MalufAlmeida}. Optical and thermodynamic properties of bumblebee AdS black holes have also been considered recently \cite{panah2026optical}. The background adopted here is the static, zero-angular-momentum sector of the BTZ-like Einstein--bumblebee solution reported by Ding \textit{et al.} \cite{Ding}; its field and potential assumptions are those specified in that work. The broader standard-model extension and its gravitational applications, together with preferred-frame vector models and Lorentz-violating electrodynamics, provide additional context for interpreting such backgrounds \cite{ColladayKostelecky,Kostelecky2004,KosteleckyTasson,JacobsonMattingly,CarrollFieldJackiw}.

A geometric extension of an optical exterior should be distinguished from a gravitational wormhole solution. Wormhole theory addresses both the geometry of throats and the matter or field content required to support them, beginning with classic constructions and extending to geometric throat criteria, energy conditions, and quantum restrictions \cite{MorrisThorne,Ellis,Bronnikov,Visser1989,HochbergVisser1997,HochbergVisser1998,FordRoman,Lobo,VisserBook,HawkingEllis}. For an optical metric, the issue is especially important because the optical exterior and any two-ended completion need not represent the same global spacetime. The intrinsic geometry, its completeness, and the range over which a Euclidean embedding exists are separate questions. Null propagation and deflection can be treated with gravitational-lensing methods, including Gauss--Bonnet formulations and finite-distance observables \cite{Perlick,GibbonsWerner,Werner,Bozza,Ishihara}; curvature and embedding are governed by standard differential geometry \cite{DoCarmo,ONeill}.

External electromagnetic fields provide another probe of curved spatial geometries. Surface-based magnetic configurations can be specified through the induced geometry and classical electrodynamics \cite{Jackson,Carter}, but such prescriptions do not by themselves constitute solutions of a coupled gravitational and electromagnetic system. In graphene, the low-energy electronic sector is described by a continuum Dirac theory, and curved or wormhole-like geometries have been used to study the influence of spatial structure and magnetic flux on quasiparticle dynamics \cite{GrapheneWormhole,RojjanasonBurikhamPimsamarn,NaderiHasanirokh2024}. A continuum treatment is distinct from a device-level model: lattice geometry, contacts, microscopic parameters, and boundary conditions are needed to connect it to transport measurements. Tight-binding analyses provide a route for assessing the validity and limits of the continuum approximation \cite{RobertsWiseman2024}. These considerations motivate studying Dirac propagation on a prescribed optical geometry while keeping the geometric model, external fields, and material realization conceptually separate.

We investigate the optical geometry of the Lorentz-violating BTZ-like background and use its prescribed two-ended completion to study null propagation and continuum Dirac scattering. The analysis distinguishes the original black-hole exterior from the optical completion and treats electromagnetic configurations as external test-field prescriptions; it does not assume a new gravitational solution or a device-specific graphene realization. Section~\ref{sec:opt} derives the optical coordinate and metric, and Sec.~\ref{sec:curv} examines intrinsic curvature. Section~\ref{sec:double} defines the reflection double; Secs.~\ref{sec:geo} and \ref{sec:embed} address null-ray access and Euclidean embedding, and Sec.~\ref{sec:em} specifies the magnetic fields. Section~\ref{sec:lit} positions the continuum model in the relevant literature, while Sec.~\ref{subsec:graphene-scattering} develops the Dirac-scattering analysis. Sections~\ref{sec:btz} and \ref{sec:scale} discuss the BTZ limit and parameter dependence; Sec.~\ref{sec:beyond-optical} considers complementary Lorentzian probes, and Sec.~\ref{sec:conc} concludes.

\section{Optical geometry}\label{sec:opt}

The optical metric provides a geometric description of the spatial
paths followed by null rays in a static spacetime. It is obtained by
conformally rescaling the spacetime metric by the inverse lapse, a
construction that converts the null condition into a unit-speed
condition on an associated Riemannian spatial metric. We use the
optical-coordinate and surface-of-revolution framework of
Refs.~\cite{GibbonsWarnick,GibbonsHerdeiro,CveticGibbons}, keeping
the radial Lorentz-violating factor in the metric rather than
absorbing it into an unexamined redefinition of the radial coordinate.
The analysis below concerns the static sector; its relation to the
stationary construction is discussed in Sec.~\ref{sec:btz}. The
distinction between the effective cosmological parameter and the
Lorentz-violating radial factor is maintained throughout: the
background relation in Eq.~(\ref{eq:2:1}) is imposed whenever a
parameter is varied, and each comparative statement specifies which
quantities are held fixed.

We use signature $(-++)$ and set $c=\hbar=1$. The static BTZ-like
line element is
\cite{Ding}
\begin{equation}
\begin{split}
 ds^2
 &=-f(r)\,\mathrm{d} t^2
   +\frac{\lambda}{f(r)}\,\mathrm{d} r^2
   +r^2\,\mathrm{d}\phi^2,\\
 f(r)
 &=-\mu-\lambda\Lambda_e r^2
   =\frac{r^2}{\ell^2}-\mu,\\
 \lambda
 &=1+\ell_{\mathrm{LV}}>0,
 \qquad
 \ell^{-2}=-\lambda\Lambda_e.
\end{split}
\label{eq:2:1}
\end{equation}
Here $\ell$ denotes the effective AdS curvature radius, $\mu>0$
is the dimensionless mass parameter, and $\phi$ is identified modulo
$2\pi$. The AdS branch adopted in this work has
$\Lambda_e<0$ and $\lambda>0$.

The horizon is located at
\begin{equation}
 r_h=\ell\sqrt{\mu}
     =\sqrt{-\frac{\mu}{\lambda\Lambda_e}},
 \qquad
 f(r_h)=0.
\label{eq:2:2}
\end{equation}
We introduce the positive optical scale
\begin{equation}
 q\equiv\frac{r_h}{\sqrt{\lambda}\,\ell^2}
   =\sqrt{-\mu\Lambda_e},
 \qquad
 \chi\equiv q\ell
   =\sqrt{\frac{\mu}{\lambda}}.
\label{eq:qscale}
\end{equation}
The relations in Eq.~(\ref{eq:qscale}) are consequences of
$r_h^2=\mu\ell^2$ and $\ell^{-2}=-\lambda\Lambda_e$;
therefore $q$ is a derived optical scale, not an independent
parameter. We take the positive square-root branches, so $q>0$.

We take Eq.~(\ref{eq:2:1}) as the static, zero-angular-momentum
sector of the BTZ-like Einstein--bumblebee solution reported in
Ref.~\cite{Ding}, under its purely radial bumblebee ansatz and the
linear Lagrange-multiplier potential used to obtain the nonzero-
cosmological-constant branch. In that solution the effective
cosmological parameter is related to the cosmological constant and
the multiplier; it is not an additional freely varying parameter
once the underlying action and potential are fixed. The present
analysis treats $\Lambda_e$ as the parameter appearing in the
specified metric and does not vary it independently of $\lambda$
when a parameter protocol requires the background relation
$\ell^{-2}=-\lambda\Lambda_e$. We do not rederive the bumblebee
field equations or independently establish stability of this
background. Our notation follows Ref.~\cite{Ding}, with $\mu$
corresponding to its mass parameter $m$.

We work on the exterior $r>r_h$, where $f(r)>0$. The condition
$\lambda>0$, or equivalently $\ell_{\mathrm{LV}}>-1$, makes
$g_{rr}=\lambda/f(r)$ positive and ensures that the optical radial
coordinate introduced below is real. No assumption
$\ell_{\mathrm{LV}}\geq0$ is needed. At $\lambda=0$ the metric
is degenerate. For $\lambda<0$ with $\Lambda_e<0$, one has
$f(r)=-\mu-\lambda\Lambda_e r^2<0$ for $r>0$; consequently,
the static exterior with $f>0$ does not exist in this parameter
branch. This is a domain restriction for the particular background
and optical construction considered here, not a model-independent
sign theorem for bumblebee couplings. Any additional restrictions
must be established from the action and its stability conditions.

For a static metric
\begin{equation}
 ds^2=-V(x)\,\mathrm{d}t^2+g_{ij}(x)\,\mathrm{d}x^i\mathrm{d}x^j,
\end{equation}
the associated optical metric is defined by
$\mathrm{d}\hat s^2=V^{-1}ds^2$
\cite{Wald,GibbonsWarnick,GibbonsHerdeiro,CveticGibbons}.
Because a conformal rescaling preserves null curves as
unparameterized curves, this construction leaves their spatial
trajectories unchanged while setting the optical lapse to unity.
In the static limit identified in Sec.~\ref{sec:btz}, the shift
vanishes and no optical wind term is present. With $V=f(r)$, one
obtains
\begin{equation}
 \mathrm{d}\hat s^2
 =-\mathrm{d} t^2+\frac{\lambda}{f(r)^2}\,\mathrm{d} r^2
  +\frac{r^2}{f(r)}\,\mathrm{d}\phi^2.
\label{eq:2:3}
\end{equation}

To put the radial part of the optical metric in proper-distance
form, introduce $\rho$ with orientation chosen such that the
conformal boundary lies at $\rho\to0^+$ and the horizon at
$\rho\to+\infty$:
\begin{equation}
 \mathrm{d}\rho=-\frac{\sqrt{\lambda}}{f(r)}\,\mathrm{d} r.
\label{eq:2:4}
\end{equation}
Introduce the dimensionless radius $u=r/r_h$. Since
$f(r)=\mu(u^2-1)$, Eq.~(\ref{eq:2:4}) becomes
\begin{equation}
 \mathrm{d}\rho=-\frac{1}{q}\frac{\mathrm{d} u}{u^2-1}.
\end{equation}
Integrating with
\begin{equation}
 \int\frac{\mathrm{d}u}{u^2-1}
 =\frac12\ln\left|\frac{u-1}{u+1}\right|
\end{equation}
and fixing the additive constant by the boundary condition
$\rho\to0$ as $u\to\infty$ gives
\begin{equation}
 \rho
 =\frac{1}{2q}\ln\left(\frac{u+1}{u-1}\right)
 =\frac{1}{q}\operatorname{arctanh}\frac{1}{u}.
\label{eq:2:5}
\end{equation}
On the exterior branch $u>1$, the logarithm is real and the map
is monotonic. It maps the entire exterior interval
$r_h<r<\infty$ bijectively onto $0<\rho<\infty$, with the two limiting
endpoints mapped as specified above.

The inverse transformation is obtained directly from
Eq.~(\ref{eq:2:5}):
\begin{equation}
 u=\coth(q\rho),
 \qquad
 r(\rho)=r_h\coth(q\rho).
\end{equation}
Under this transformation, the angular coefficient becomes
\begin{equation}
 \frac{r^2}{f(r)}
 =\ell^2\frac{u^2}{u^2-1}
 =\ell^2\cosh^2(q\rho).
\end{equation}
Defining the optical circumferential-radius function by
\begin{equation}
 C(\rho)\equiv\sqrt{\frac{r^2}{f(r)}},
\end{equation}
we find the exact expressions
\begin{equation}
 r(\rho)=r_h\coth(q\rho),
 \qquad
 C(\rho)=\ell\cosh(q\rho).
\label{eq:2:6}
\end{equation}
The quantity $C(\rho)$ is the circumferential radius measured in
the optical spatial geometry: a circle at fixed $\rho$ has length
$2\pi C(\rho)$. It must not be identified with the original
areal coordinate $r$. By construction, the radial optical line
element is $\mathrm{d}\rho^2$, and the angular part is
$C(\rho)^2\mathrm{d}\phi^2$.

Combining the radial and angular transformations yields the exact
ultrastatic optical spacetime
\begin{equation}
 \mathrm{d}\hat s^2
 =-\mathrm{d} t^2+\mathrm{d}\rho^2+C(\rho)^2\,\mathrm{d}\phi^2.
\label{eq:2:10}
\end{equation}
The metric is the product $\mathbb{R}\times\Sigma$, with
optical spatial section $\Sigma$ described by
\begin{equation}
 \mathrm{d}\hat s_{\mathrm{sp}}^2
 =\mathrm{d}\rho^2+C(\rho)^2\,\mathrm{d}\phi^2,
 \qquad 0<\rho<\infty.
\end{equation}
With $\phi\sim\phi+2\pi$, the original exterior optical section has
the topology $(0,\infty)\times S^1$, with the conformal-boundary
end at $\rho\to0^+$ and the horizon at infinite optical distance,
$\rho\to+\infty$. We determine its intrinsic curvature in
Sec.~\ref{sec:curv}. Parameter comparisons at fixed $\ell$ and fixed
$\Lambda_e$ are kept distinct in Sec.~\ref{sec:scale}, since the
relation $\ell^{-2}=-\lambda\Lambda_e$ couples these quantities
when $\lambda$ varies.

\section{Intrinsic curvature}\label{sec:curv}

With the optical metric established, we now characterize its
intrinsic curvature independently of any embedding construction.
An embedding diagram can visualize a metric, but it does not define
its curvature; the latter is fixed entirely by the intrinsic line
element. For the rotationally symmetric metric obtained above,
the curvature follows directly from the circumferential-radius
function $C(\rho)$, using the standard method employed in BTZ
optical geometry
\cite{GibbonsWarnick,GibbonsHerdeiro,CveticGibbons}.

For a two-dimensional rotationally symmetric metric of the form
\begin{equation}
 \mathrm{d}\sigma^2=\mathrm{d}\rho^2+C(\rho)^2\,\mathrm{d}\phi^2,
\end{equation}
the Gaussian curvature is \cite{CveticGibbons}
\begin{equation}
 \mathcal{K}=-\frac{C''(\rho)}{C(\rho)}.
\end{equation}
Consequently, the Gaussian curvature is constant and negative \cite{CveticGibbons}:
\begin{equation}
\begin{aligned}
 \mathcal{K}_{\mathrm{opt}}
 &=-\frac{C''(\rho)}{C(\rho)}
 =-q^2=\mu\Lambda_e
 =-\frac{\mu}{\lambda\ell^2},\\
 \mathcal{R}_{\mathrm{opt}}
 &=2\mathcal{K}_{\mathrm{opt}}=-2q^2.
\end{aligned}
\label{eq:3:1}
\end{equation}
The symbols $\mathcal{K}_{\mathrm{opt}}$ and
$\mathcal{R}_{\mathrm{opt}}$ refer exclusively to the Gaussian
and scalar curvatures of the two-dimensional optical spatial
metric. 

In two dimensions the Riemann tensor is determined by the Gaussian
curvature, so the optical spatial Kretschmann invariant and curvature
radius follow directly:
\begin{equation}
R_{ijkl}R^{ijkl}=4q^4,\qquad
\mathcal L_{\mathrm{opt}}=q^{-1}.
\label{eq:3:2}
\end{equation}
The optical spatial section is therefore locally hyperbolic with
curvature radius $q^{-1}$. Globally, however, it is the quotient
with angular identification $\phi\sim\phi+2\pi$ restricted
to the exterior range $\rho>0$; it is not the complete
hyperbolic plane.

The Lorentz-violating dependence of the curvature must be assessed
with the background constraint $\ell^{-2}=-\lambda\Lambda_e$
explicitly imposed. If $\ell$ and $\mu$ are held fixed, then
$\mathcal{K}_{\mathrm{opt}}=-\mu/(\lambda\ell^2)$ scales as
$\lambda^{-1}$. If instead $\Lambda_e$ and $\mu$ are fixed,
then $\ell$ varies with $\lambda$ and
$\mathcal{K}_{\mathrm{opt}}=\mu\Lambda_e$ is independent of
$\lambda$. These comparisons describe different parameter
families and are distinguished in Sec.~\ref{sec:scale}.

The horizon lies at infinite optical radial distance. This follows
explicitly by introducing the near-horizon variable
\begin{equation}
 \epsilon\equiv\frac{r-r_h}{r_h}\ll1,
 \qquad u=1+\epsilon.
\end{equation}
Equation~(\ref{eq:2:5}) then gives
\begin{equation}
 \rho
 =\frac{1}{2q}\ln\left(\frac{2+\epsilon}{\epsilon}\right)
 \simeq\frac{1}{2q}\ln\left(\frac{2}{\epsilon}\right).
\label{eq:3:3}
\end{equation}
The horizon is therefore an end at infinite optical radial
distance. This distinguishes it from the finite-coordinate
conformal-boundary end at $\rho=0$, where the doubling described
in the next section is performed.

\section{The two-ended optical wormhole completion}\label{sec:double}

The coordinate map of Sec.~\ref{sec:opt} covers only the original
black-hole exterior, $\rho>0$: the limit $\rho\to0^+$ is the
conformal boundary $r\to\infty$, and $\rho\to+\infty$ is the
horizon. Although the optical profile
$C(\rho)=\ell\cosh(q\rho)$ is smooth and even, extending this
function to negative $\rho$ does not extend the original BTZ radial
coordinate chart. In particular, the negative-$\rho$ branch is not
an additional exterior of the original black hole.

To construct a two-ended completion, we therefore take two copies
of the optical half-cylinder and identify their boundary circles
at $\rho=0$. The signed coordinate $\rho$ is then a global
Gaussian normal coordinate on the doubled manifold, whose metric
is
\begin{equation}
\begin{aligned}
 \mathrm{d}\hat s_{\mathrm{wh}}^2
 &=-\mathrm{d} t^2+\mathrm{d}\rho^2+C(\rho)^2\,\mathrm{d}\phi^2,\\
 C(\rho)&=\ell\cosh(q\rho),
 \qquad \rho\in\mathbb{R}.
\end{aligned}
\label{eq:4:1}
\end{equation}
The even profile is smooth at the joining surface and satisfies
\begin{equation}
 C'(0)=0,\qquad C''(0)=q^2\ell>0.
\end{equation}
The metric coefficients are regular at the joining circle.
Moreover, the same smooth metric function describes both sides,
so the reflection introduces neither a jump in the first
fundamental form nor a distributional curvature layer at
$\rho=0$. Since $C(\rho)$ has a unique global minimum there,
the joining circle is the throat, with radius and circumference
\begin{equation}
 C_{\mathrm{th}}=\ell,\qquad
 \mathcal{C}_{\mathrm{th}}=2\pi\ell.
\end{equation}
This is a regular minimal-circumference throat of the intrinsic
two-ended geometry, in the geometric sense of the throat criterion
of Morris and Thorne \cite{MorrisThorne}. The construction is
well-defined for every $\mu>0$ and $\lambda>0$. Its smoothness and
completeness are statements about the prescribed ultrastatic
optical metric. They do not establish geodesic completeness of the
original Lorentzian black-hole exterior, nor do they supply a global
bumblebee-vector configuration or boundary data for a gravitational
solution.

The doubled manifold has two asymptotic ends, one at each of
$\rho\to+\infty$ and $\rho\to-\infty$. At both ends,
\begin{equation}
 C(\rho)\sim\frac{\ell}{2}e^{q|\rho|},
 \qquad
 \mathcal{K}_{\mathrm{opt}}=-\frac{C''}{C}=-q^2.
\label{eq:4:2}
\end{equation}
The circumference grows exponentially toward either end, while
the Gaussian curvature remains $-q^2$. The proper radial distance
from the throat to either asymptotic end is
\begin{equation}
 \int_0^\infty\mathrm{d}\rho=\infty.
\end{equation}
The spatial metric is complete. Indeed, $\rho$ ranges over the
entire real line and the compact angular factor has radius bounded
below by $\ell>0$; consequently, a finite-length curve cannot
escape to either $|\rho|\to\infty$ end, and the angular factor
introduces no incompleteness. The ends are asymptotically
hyperbolic in the sense of Eq.~(\ref{eq:4:2}), not asymptotically
flat.

Because the doubled metric is ultrastatic, its lapse is unity
and the Killing field $\partial_t$ is everywhere timelike;
there is no Killing horizon in this constructed spacetime. A
radial null curve obeys $\mathrm{d}t=\pm\mathrm{d}\rho$ and
therefore crosses the throat in finite coordinate time and affine
parameter. The doubled metric is traversable in this geometric
sense. This property does not establish that the completion
solves the original Einstein--bumblebee equations or that its
physical source obeys a specified energy condition.

The throat's flare-out behavior can also be characterized in
areal-radius coordinates, without invoking an embedding diagram.
Set $R=C(\rho)$ on either branch. The exact profile
$R=\ell\cosh(q\rho)$ gives
\begin{equation}
 \left(\frac{\mathrm{d} R}{\mathrm{d}\rho}\right)^2
 =q^2(R^2-\ell^2).
\end{equation}
For $R>\ell$, where the areal-radius coordinate is locally
invertible on each branch, the spatial metric takes the form
\begin{equation}
 \mathrm{d}\sigma^2
 =\frac{\mathrm{d} R^2}{1-b_{\mathrm{sh}}(R)/R}
  +R^2\,\mathrm{d}\phi^2,
\end{equation}
where the shape function is
\begin{equation}
 b_{\mathrm{sh}}(R)
 =R\left[1-q^2(R^2-\ell^2)\right].
\label{eq:4:3}
\end{equation}
At $R=\ell$, one has
\begin{equation}
 b_{\mathrm{sh}}(\ell)=\ell,
 \qquad
 b_{\mathrm{sh}}'(\ell)=1-2q^2\ell^2<1
 \quad(q>0),
\end{equation}
This satisfies the standard local flare-out inequality, adapted
to the present $2+1$-dimensional spatial geometry. The
coordinate $R$ is two-to-one on the doubled manifold and fails
to be a regular coordinate at the throat; the signed proper
coordinate $\rho$ is the global regular coordinate and must be
used to cover both branches.

The geometric construction must now be distinguished from a
solution of the underlying gravitational theory. The ultrastatic
product structure, together with the spatial curvature in
Eq.~(\ref{eq:3:1}), fixes the Ricci tensor of the doubled metric:
\begin{equation}
 R_{tt}=0,\qquad
 R_{ij}=-q^2h_{ij},\qquad
 R=-2q^2,
\end{equation}
where $h_{ij}$ is the two-dimensional spatial metric. With the
Einstein-tensor convention
$G_{\mu\nu}=R_{\mu\nu}-\tfrac12Rg_{\mu\nu}$, this gives
\begin{equation}
 G_{tt}=-q^2,\qquad G_{ij}=0.
\end{equation}
To identify the source required by a particular comparison theory, we assume, solely for this diagnostic purpose, that the metric satisfies the standard $(2+1)$-dimensional Einstein equations with vanishing cosmological constant:
\begin{equation}
 G_{\mu\nu}=8\pi G_3T_{\mu\nu}^{\mathrm{eff}}.
\end{equation}
Under this additional assumption, the inferred effective source
has the following orthonormal-frame components:
\begin{equation}
 \rho_{\mathrm{eff}}=-\frac{q^2}{8\pi G_3},
 \qquad
 p_{\hat\rho}=p_{\hat\phi}=0,
 \qquad G_3>0.
\label{eq:4:4}
\end{equation}
For a radial null vector with orthonormal components
$k^{\hat a}=(1,1,0)$, it follows that
\begin{equation}
 T_{\mu\nu}^{\mathrm{eff}}k^\mu k^\nu
 =\rho_{\mathrm{eff}}+p_{\hat\rho}
 =-\frac{q^2}{8\pi G_3}<0.
\end{equation}
The effective source in this ordinary-Einstein interpretation
therefore violates the null energy condition. The conclusion is
strictly conditional on the assumed Einstein equation with zero
cosmological constant; it is not an energy-condition statement
about the bumblebee field or any matter sector of the original
gravity model.

Neither the optical conformal rescaling nor the reflection
identification is shown to preserve the Einstein--bumblebee field
equations. Establishing a self-consistent traversable wormhole
solution in that theory would require solving the coupled
gravitational and bumblebee equations with a globally defined
field configuration and appropriate boundary conditions.

The relation between the original exterior and its completion
is thus precise: the $\rho>0$ branch of Eq.~(\ref{eq:4:1})
reproduces the original optical spatial metric, whereas the
$\rho<0$ branch is added by reflection and has no independent
BTZ radial-coordinate interpretation. The inverse map
$r(\rho)=r_h\coth(q\rho)$ applies only for $\rho>0$ and
cannot be interpreted as a second physical black-hole exterior.
Accordingly, Eq.~(\ref{eq:4:1}) defines a complete two-ended
optical completion, not a reinterpretation of the original
black-hole spacetime without modification. This distinction
between an effective geometry and its physical realization is
also central to optical-wormhole and graphene analogues
\cite{GrapheneWormhole}.

\section{Null geodesics and throat accessibility}\label{sec:geo}

We next determine which null geodesics of the doubled optical
spacetime can reach or cross the throat. In an ultrastatic metric,
the spatial projection of a null geodesic is a geodesic of the
optical spatial metric, and the increment in coordinate time equals the optical arclength
along the spatial projection. For an affinely parametrized null geodesic
with parameter $s$, time-translation and rotational symmetry
provide the conserved energy (\(\mathcal E\)) and angular momentum (\(\mathcal L\))
\cite{OpticalWormhole}:
\begin{equation}
\mathcal E=\dot t,\qquad \mathcal L=C(\rho)^2\dot\phi,\qquad
\dot\rho^2+\frac{\mathcal L^2}{C(\rho)^2}=\mathcal E^2,
\label{eq:5:1}
\end{equation}
where a dot denotes differentiation with respect to $s$. For
future-directed rays, take $\mathcal E>0$ and introduce the
impact parameter $b=\mathcal L/\mathcal E$ together with the
dimensionless magnitude $\beta=|b|/\ell$. The radial equation
then contains the effective centrifugal term
\begin{equation}
V_{\mathrm{eff}}(\rho)=\frac{\mathcal L^2}{\ell^2}\sech^2(q\rho)
\label{eq:5:2}
\end{equation}
which reaches its unique maximum at $\rho=0$, where
$C(0)=\ell$. A ray crosses the throat precisely when its squared energy exceeds
the maximum of this barrier, equivalently
\begin{equation}
|b|<\ell\quad(\beta<1).
\label{eq:5:3}
\end{equation}

At the critical impact parameter $|b|=\ell$, the radial
velocity vanishes at the throat. Expanding
$\sech^2(q\rho)=1-q^2\rho^2+O(\rho^4)$ gives
$\dot\rho^2\simeq\mathcal E^2q^2\rho^2$ near $\rho=0$;
the critical orbit therefore approaches the throat only
asymptotically in affine parameter and forms the separatrix.
For $|b|>\ell$, the ray encounters a turning point determined by
\begin{equation}
\cosh(q\rho_{\mathrm{turn}})=\frac{|b|}{\ell},
\qquad
|\rho_{\mathrm{turn}}|=\frac{1}{q}\operatorname{arcosh}\beta,
\label{eq:5:4}
\end{equation}
and is reflected on its initial side, without reaching the throat.

The total angular sweep along each allowed trajectory follows
from the first integrals:
\begin{equation}
\left|\frac{d\phi}{d\rho}\right|=\frac{|b|}{C(\rho)^2\sqrt{1-b^2/C(\rho)^2}}.
\label{eq:5:5}
\end{equation}
For a throat-crossing ray, the substitution
$y=\tanh(q\rho)$ maps the two asymptotic ends to $y=\pm1$
and reduces the angular integral to
$\frac{|b|}{q\ell^2}\int_{-1}^{1}
\frac{\mathrm{d}y}{\sqrt{1-\beta^2+\beta^2y^2}}$.
Its exact evaluation yields
\begin{equation}
|\Delta\phi|_{\mathrm{cross}}
=\frac{2}{q\ell}\operatorname{artanh}\beta,
\qquad 0\leq\beta<1.
\label{eq:5:6}
\end{equation}
For a returning ray on one side, the turning point corresponds
to $y_{\mathrm{turn}}=\sqrt{1-\beta^{-2}}$. The total sweep
is twice the integral from this turning point to the asymptotic
end $y=1$; evaluating it gives
\begin{equation}
|\Delta\phi|_{\mathrm{return}}
=\frac{2}{q\ell}\operatorname{artanh}\frac{1}{\beta},
\qquad \beta>1.
\label{eq:5:7}
\end{equation}
Equations~(\ref{eq:5:6}) and~(\ref{eq:5:7}) give the total
angular change along the complete crossing or return trajectory.
They are not deflection angles defined relative to an independently
chosen reference path. Both branches diverge logarithmically as
$\beta\to1^-$ or $\beta\to1^+$, respectively, as expected
for trajectories approaching the critical throat separatrix.
Furthermore, since
\begin{equation}
q\ell=\chi,
\label{eq:5:8}
\end{equation}
the parameter $\chi=q\ell=\sqrt{\mu/\lambda}$ sets the
angular scale of these trajectories, whereas the critical impact
parameter remains $|b|=\ell$. This separates the Lorentz-violating
effect on ray accumulation from the geometric threshold for throat
crossing and permits a direct comparison with the undeformed
optical-wormhole problem \cite{BTZWormhole,OpticalWormhole}.
The exact formulas admit independent limiting checks. At \(b=0\), the
crossing ray is radial and Eq.~(\ref{eq:5:6}) gives zero angular sweep.
For \(\beta\ll1\),
\[
|\Delta\phi|_{\rm cross}=\frac{2}{q\ell}
\left(\beta+\frac{\beta^3}{3}+O(\beta^5)\right),
\]
which is regular. For \(\beta\gg1\), the returning sweep is
\[
|\Delta\phi|_{\rm return}=\frac{2}{q\ell}
\left(\beta^{-1}+\frac{\beta^{-3}}{3}+O(\beta^{-5})\right),
\]
consistent with a weakly bending, nearly straight trajectory. Near the
separatrix, both branches diverge logarithmically,
\[
|\Delta\phi|\sim-\frac{1}{q\ell}\ln|1-\beta|+O(1),
\]
with branch-dependent finite terms. These limits check the normalization,
branch assignment, and critical behavior of the exact integrals. The
angular sweep is not a conventional asymptotic deflection angle: that
quantity would require a reference geometry and endpoint prescription.

\section{Euclidean embedding}\label{sec:embed}

The intrinsic geometry and its representation as a surface in
Euclidean three-space are separate issues. Completeness and smooth
negative curvature do not guarantee a global isometric embedding
in ordinary Euclidean space. For a rotationally symmetric metric,
the restriction is encoded in the meridional embedding equation;
a finite embedding range is already familiar in the BTZ optical
geometry \cite{GibbonsWarnick,GibbonsHerdeiro,CveticGibbons}. We now derive the corresponding
endpoint for the Lorentz-violating optical profile.
Embed the spatial optical surface as a surface of revolution in Euclidean $\mathbb E^3$:
\begin{equation}
 X=C(\rho)\cos\phi,\quad Y=C(\rho)\sin\phi,\quad Z=Z(\rho).
 \label{eq:6:1}
\end{equation}
The induced Euclidean metric is
$(C'^2+Z'^2)\mathrm{d}\rho^2+C^2\mathrm{d}\phi^2$.
Matching it to the intrinsic optical metric gives
\begin{equation}
 \left(\frac{\dd Z}{\dd\rho}\right)^2
 =1-\left(\frac{\dd C}{\dd\rho}\right)^2
 =1-\ell^2q^2\sinh^2(q\rho).
 \label{eq:6:2}
\end{equation}
The embedding profile is therefore
\begin{equation}
 Z(\rho)=Z_0\pm\int_0^\rho
 \sqrt{1-\ell^2q^2\sinh^2(q\bar\rho)}\,\dd\bar\rho.
 \label{eq:6:3}
\end{equation}
On the doubled manifold this condition is symmetric under
$\rho\mapsto-\rho$. Reality of the Euclidean embedding
requires
\begin{equation}
\begin{aligned}
 |\sinh(q\rho)|\leq\frac{1}{\ell q}&=\frac{1}{\ell\sqrt{-\mu\Lambda_e}} =\sqrt{\frac{\lambda}{\mu}},\\
 |\rho|\leq\rho_{\max}
 &=\frac{1}{\sqrt{-\mu\Lambda_e}}\operatorname{arsinh}\sqrt{\frac{\lambda}{\mu}}.
\end{aligned}
 \label{eq:6:4}
\end{equation}
On the original exterior branch $\rho>0$, the same condition becomes
\begin{equation}
 r^2\geq\mu\ell^2\left(1+\frac{\mu}{\lambda}\right).
 \label{eq:6:5}
\end{equation}
The endpoint condition has a direct geometric meaning. For a
surface of revolution with profile radius $C(\rho)$, the
meridional metric is $(C'^2+Z'^2)\mathrm{d}\rho^2$; hence an
isometric Euclidean embedding requires $C'^2\leq1$. In the
present case $C'=\ell q\sinh(q\rho)$, whose magnitude grows
without bound toward either asymptotic end. At
$|\rho|=\rho_{\max}$, one has $Z'=0$ and $|C'|=1$, so the
meridional tangent is horizontal in the $(C,Z)$ profile plane.
For larger $|\rho|$, the quantity $1-C'^2$ is negative and no
real profile function $Z(\rho)$ exists within this surface-of-
revolution ansatz. Since $\rho_{\max}$ is finite for every
finite $\lambda>0$ while the horizon lies at $\rho=\infty$,
the Euclidean embedding terminates before reaching the horizon.
This is an embedding obstruction, not a curvature singularity or
incompleteness of the optical metric. The Lorentz-violating
parameter shifts the endpoint but cannot eliminate the finite
range for finite positive $\lambda$. The endpoint is controlled by
\(\chi=q\ell=\sqrt{\mu/\lambda}\):
\[
\frac{\rho_{\max}}{\ell}=\frac{1}{\chi}
\operatorname{arsinh}\frac{1}{\chi}.
\]
It is therefore unchanged by parameter variations preserving $\chi$,
even if the dimensional curvature scale changes. The embedding profile
is symmetric under $\rho\mapsto-\rho$ together with reflection
$Z-Z_0\mapsto-(Z-Z_0)$; equivalently, $Z(\rho)-Z_0$ is odd,
whereas $Z'(\rho)$ is even. At both endpoints $Z'=0$, and beyond them no real
profile exists within this Euclidean surface-of-revolution ansatz.
This obstruction does not rule out embeddings in a different ambient
space or embedding class.

\section{Electromagnetic fields}\label{sec:em}

The optical geometry admits test electromagnetic configurations, but the metric does not determine an electromagnetic field: the latter must be specified independently. We consider two axisymmetric magnetic prescriptions that are often distinguished in surface-based constructions: a field whose direction is fixed along the symmetry axis of the Euclidean embedding, and a field whose component normal to the surface is constant \cite{DoCarmo,Jackson}. These prescriptions coincide only when the normal projection of the uniform axial field is constant, as for a planar surface with fixed normal orientation. Here the meridional normal varies with \(\rho\), so the resulting gauge potentials and field profiles are inequivalent. The construction is geometric and is not a solution of Maxwell equations coupled to the bumblebee gravitational sector.

Let the surface gauge potential be
\begin{equation}
 A=A_\phi(C)\,\dd\phi,\qquad F=\dd A
 =\frac{\dd A_\phi}{\dd C}\,\dd C\wedge\dd\phi.
 \label{eq:7:1}
\end{equation}
All embedding-based field expressions below are defined only on the Euclidean-embeddable portion of the doubled optical surface, \( |\rho|\leq\rho_{\max}\), with \(\rho_{\max}\) given in Eq.~(\ref{eq:6:4}). The intrinsic doubled metric itself is defined for every \(\rho\in\mathbb R\); the finite interval is a limitation of its realization as a surface of revolution in \(\mathbb E^3\), not a boundary of the optical manifold. Expressions written in terms of the original coordinate \(r\) refer exclusively to the \(\rho>0\) exterior branch.

For an axisymmetric surface, the axial field \(B_Z\) and the signed normal component \(B_n\) are related to the gauge potential by
\begin{equation}
\begin{aligned}
 B_Z(C)&=\frac{1}{C}\frac{\dd A_\phi}{\dd C},\\
 B_n(\rho)&=\frac{1}{C}\frac{\dd A_\phi}{\dd C}
 \frac{\dd C}{\dd\rho}
 =B_Z(C)\,C'(\rho).
\end{aligned}
\label{eq:7:2}
\end{equation}
The sign of \(B_n\) is fixed by the chosen orientation of the surface normal; with the global \(\rho\)-orientation, \(C'(\rho)\) changes sign across the throat.

\paragraph{Uniform axial field.}
For a constant field \(B_Z=B_0\) parallel to the embedding-space \(Z\)-axis, integration of Eq.~(\ref{eq:7:2}) gives
\begin{equation}
\begin{split}
 A_\phi(C)&=\frac{B_0}{2}C^2+A_0,\\
 B_n(\rho)&=B_0\ell q\sinh(q\rho),\\
 A_\phi(r)&=\frac{B_0r^2}{2f(r)}+A_0.
\end{split}
\label{eq:7:3}
\end{equation}
Thus the normal component vanishes at the throat and reverses sign between the two asymptotic ends, even though the ambient axial field is uniform. This behavior reflects the change in orientation of the surface normal relative to the fixed \(Z\)-axis.

\paragraph{Uniform normal field.}
If instead the signed normal component is prescribed to be \(B_n=B_c\), Eq.~(\ref{eq:7:2}) requires
\begin{equation}
\begin{split}
 \frac{\dd A_\phi}{\dd\rho}&=B_cC(\rho),\\
 A_\phi(\rho)&=\frac{B_c\ell}{q}\sinh(q\rho)+A_0,\\
 A_\phi(r)&=\frac{B_c\sqrt{\lambda}\ell^2}{\sqrt{f(r)}}+A_0.
\end{split}
\label{eq:7:4}
\end{equation}
The corresponding ambient axial field is not uniform:
\begin{equation}
 B_Z(\rho)=\frac{B_c}{\ell q\sinh(q\rho)}.
 \label{eq:7:5}
\end{equation}
Since \(C'(0)=0\), maintaining a nonzero constant normal component requires \(B_Z\) to diverge at the throat. The divergence belongs to this imposed field prescription; it is not an intrinsic geometric singularity. In particular, the uniform-normal-field gauge potential is smooth as a function of \(\rho\) at the throat, whereas the ambient axial field needed to realize it becomes singular there. The two magnetic configurations in Eqs.~(\ref{eq:7:3})--(\ref{eq:7:5}) should therefore not be conflated.

Equations~(\ref{eq:7:1})--(\ref{eq:7:5}) define a test one-form and its pullback field strength on the embeddable patch. They do not incorporate electromagnetic backreaction, a bumblebee-vector coupling, or boundary conditions at the ends. In particular, no Maxwell equation with a specified current or material response has been solved; the profiles are admissible geometric prescriptions on the stated patch, not dynamically selected electromagnetic solutions. The additive constant \(A_0\) is locally removable by a gauge transformation. Globally, because \(\phi\) is periodic, its removal is subject to the admissible large gauge transformations and to the phase boundary condition of any charged probe; it can therefore encode a holonomy rather than being automatically physically irrelevant.

An electrostatic profile can be introduced independently as
\begin{equation}
 A_{\rm em}=\Phi(C)\,\dd t,\qquad
 F=\dd A_{\rm em}
 =\frac{\dd\Phi}{\dd C}\frac{\dd C}{\dd\rho}
 \,\dd\rho\wedge\dd t.
 \label{eq:7:6}
\end{equation}
The function \(\Phi(C)\) is external input. Unlike stationary rotating optical geometries, the metric considered here has no rotational shift and no Zermelo wind; consequently, it does not induce the rotation-dependent effective electric connection associated with a nonzero shift vector \cite{GibbonsHerdeiro,CveticGibbons}.

\section{Relation to existing optical and magnetized-wormhole analyses}\label{sec:lit}

The undeformed reference is the static BTZ optical geometry; its
optical-metric and graphene analogue was developed in
Ref.~\cite{CveticGibbons}, building on the BTZ background
\cite{BTZ,BTZ2}. Optical-metric wormhole constructions and their
ray/wave analyses also have direct precedents~\cite{BTZWormhole,OpticalWormhole}.
Optical and thermodynamic signatures of Lorentz-symmetry breaking in
bumblebee AdS black holes have also been studied
\cite{panah2026optical}; the present analysis instead focuses on the
optical spatial geometry, its reflection completion, null propagation,
and continuum Dirac scattering. Here the black-hole exterior remains
one-sided; the second end is introduced only through the reflection
double in Eq.~(\ref{eq:4:1}). The deformation factor $\lambda$ is
retained in the optical-coordinate map, while the static limit and the
effective-mass relation are checked separately in
Secs.~\ref{sec:btz} and~\ref{sec:scale}. The background itself is
adopted from the Einstein--bumblebee solution of Ref.~\cite{Ding}; the
present work does not rederive its field equations or claim a
self-consistent gravitational wormhole.

The magnetic configurations are prescribed test fields on the
Euclidean-embeddable patch, not a coupled
Einstein--bumblebee--Maxwell solution. Their distinction follows from
the relation between the ambient axial direction and the varying
surface normal, using standard surface geometry and electrodynamics
\cite{DoCarmo,Jackson}, with direct magnetized-wormhole and curved-graphene
transport precedents provided by
Refs.~\cite{RojjanasonBurikhamPimsamarn,NaderiHasanirokh2024}. The field
profiles are therefore conditional on the chosen embedding and do not
represent dynamically selected solutions.

The Dirac sector is a continuum matter-wave problem on the doubled
intrinsic metric. Curved-graphene and wormhole transport studies
provide the relevant context~\cite{GrapheneWormhole,RojjanasonBurikhamPimsamarn,NaderiHasanirokh2024,Gallerati2019}, while
 tight-binding-to-continuum analyses assess both the validity and the
limitations of curved-space Dirac descriptions for deformed graphene
\cite{RobertsWiseman2024,RobertsWiseman2022}. The scattering amplitudes
derived below are exact for the stated ideal continuum Hamiltonian;
translating them into a device observable requires a specified lattice
deformation, valley/sublattice structure, gauge holonomy, contacts, and
boundary conditions. The relevant gauge-field and valley background
is reviewed in Refs.~\cite{Semenoff1984,Haldane1988,JackiwPi2007,GuineaKatsnelsonGeim2009,VozmedianoKatsnelsonGuinea2010}; these works provide context for related ingredients, rather than a single mechanism identical to the one considered here (see Table~\ref{tab:core-results}).

\begin{table*}[t]
\centering
\caption{Principal exact results, equation anchors, and validity domains. The table is a results summary; defining expressions are given in the cited equations.}
\label{tab:core-results}
\small
\setlength{\tabcolsep}{4pt}
\begin{ruledtabular}
\begin{tabular}{@{}p{0.20\textwidth}p{0.31\textwidth}p{0.39\textwidth}@{}}
Sector & Defining result & Domain and interpretation\\
\hline
Optical geometry & Coordinate map and metric: Eqs.~(\ref{eq:2:5})--(\ref{eq:2:10}); curvature: Eq.~(\ref{eq:3:1}) & Original exterior; intrinsic optical geometry\\
Reflection double & Metric and throat: Eq.~(\ref{eq:4:1}) & Complete geometric completion; not a sourced bumblebee solution\\
Null propagation & Threshold and angular sweeps: Eqs.~(\ref{eq:5:3})--(\ref{eq:5:7}) & Null geodesics of the doubled ultrastatic metric\\
Euclidean embedding & Reality interval: Eq.~(\ref{eq:6:4}) & Surface-of-revolution embedding only\\
Magnetic test fields & Gauge potentials and field profiles: Eqs.~(\ref{eq:7:3})--(\ref{eq:7:5}) & Prescribed fields on the embeddable patch\\
Continuum Dirac scattering & Amplitudes, probabilities, pole lattice, residues, and delay: Eqs.~(\ref{eq:gr:14})--(\ref{eq:gr:37}) & Ideal full-line continuum model; conditional packet pole expansion, not device conductance\\
Lorentzian probes & Invariants and radial operators: Eqs.~(\ref{eq:beyond-invariants})--(\ref{eq:beyond-KG-radial}) & Original Lorentzian background; candidate discriminants only\\
\end{tabular}
\end{ruledtabular}
\end{table*}

\subsection{A graphene Dirac analogue: exact scattering and resonance structure}
\label{subsec:graphene-scattering}

The optical metric supplies a well-defined continuum geometry on which to formulate massless Dirac dynamics for graphene quasiparticles. This is a condensed-matter analogue of the optical spatial geometry, not a realization of the Lorentz-violating gravitational field equations. The connection between BTZ optical geometry and graphene was established in Ref.~\cite{CveticGibbons}; the foundational electronic and transport properties of graphene are reviewed in Refs.~\cite{Novoselov2004,GeimNovoselov2007,CastroNeto2009}. Subsequent work on curved-graphene transport and continuum limits of tight-binding models provides complementary context \cite{NaderiHasanirokh2024,RobertsWiseman2024}. Here we formulate the Dirac operator on the doubled metric of Eq.~(\ref{eq:4:1}) and solve its ideal two-ended scattering problem, drawing on standard Dirac scattering and spectral methods~\cite{Novoselov2004,GeimNovoselov2007,CastroNeto2009,Thaller1992,Novikov2007,Katsnelson2006,Beenakker2008,TaylorScattering,SakuraiNapolitano2017,LandauLifshitzQM,Morse1929}. The resulting scalar partner equations belong to the Scarf-II class of hyperbolic potentials \cite{CooperKhareSukhatme1995,KhareSukhatme1988,LevaiEtAl2017,NISTDLMF}. All results in this subsection concern the continuum model; a finite device additionally requires a lattice realization, contacts, and boundary conditions.

To make the condensed-matter energy scale explicit, we restore $\hbar$ throughout this subsection and denote the Fermi velocity of graphene by $v_F$. The ultrastatic effective spacetime governing the quasiparticle dynamics is then given by
\begin{equation}
 ds_{\mathrm{gr}}^2=-v_F^2dt^2+d\rho^2+C(\rho)^2d\phi^2,
 \quad C(\rho)=\ell\cosh(q\rho).
 \label{eq:gr:1}
\end{equation}
This effective metric has the same spatial metric as Eq.~(\ref{eq:4:1}); $v_F$ is the characteristic speed of the low-energy Dirac theory. The covariant spinor formalism follows standard curved-spacetime Dirac theory~\cite{Thaller1992,Fock1929,Weyl1929,BirrellDavies1982,ParkerToms2009}. Choose the orthonormal coframe
\begin{equation}
 e^{\hat 0}=v_Fdt,\qquad e^{\hat 1}=d\rho,\qquad
 e^{\hat 2}=C(\rho)d\phi .
 \label{eq:gr:2}
\end{equation}
The spatial spin connection is evaluated below from the torsion-free
Cartan equation, using the coframe in Eq.~(\ref{eq:gr:2}).
We derive the operator rather than assume its form. The curved-space
continuum description of graphene couples the low-energy two-dimensional
Dirac equation to the induced metric
\cite{GrapheneWormhole,VozmedianoDeJuanCortijo2008,CortijoVozmediano2007}.
For the effective spacetime metric in Eq.~(\ref{eq:gr:1}), let
$\eta_{\hat a\hat b}=\mathrm{diag}(-1,1,1)$ and use the coframe in
Eq.~(\ref{eq:gr:2}). Its dual frame is
\begin{equation}
 e_{\hat 0}=v_F^{-1}\partial_t,\qquad
 e_{\hat 1}=\partial_\rho,\qquad
 e_{\hat 2}=C^{-1}\partial_\phi.
 \label{eq:gr:2a}
\end{equation}
The torsion-free Cartan equation
$de^{\hat a}+\omega^{\hat a}{}_{\hat b}\wedge e^{\hat b}=0$
gives
\begin{equation}
\omega^{\hat 2}{}_{\hat 1}=C'd\phi,\qquad
\omega^{\hat 1}{}_{\hat 2}=-C'd\phi,
\label{eq:gr:3}
\end{equation}
with $C'/C=q\tanh(q\rho)$.  We use the representation
$\gamma^{\hat 0}=i\sigma^3$,
$\gamma^{\hat 1}=\sigma^2$, and
$\gamma^{\hat 2}=\sigma^1$, which satisfies the stated Clifford algebra
and fixes the Pauli-matrix convention in Eq.~(\ref{eq:gr:4}).
With Clifford matrices satisfying
$\{\gamma^{\hat a},\gamma^{\hat b}\}=2\eta^{\hat a\hat b}$,
the spinor derivative is
\begin{equation}
 \nabla_\mu\Psi=\left(\partial_\mu+\Gamma_\mu\right)\Psi,\qquad
 \Gamma_\mu=\frac18\omega_{\mu\hat a\hat b}
 [\gamma^{\hat a},\gamma^{\hat b}],
 \label{eq:gr:2b}
\end{equation}
where the antisymmetric frame indices are summed. The covariant massless
Dirac equation is
\begin{equation}
 i\hbar v_F\gamma^{\hat a}e_{\hat a}{}^\mu\nabla_\mu\Psi=0.
 \label{eq:gr:2c}
\end{equation}
Substituting the dual frame and spin connection, and multiplying by the
time-like gamma matrix to obtain Schrödinger form, gives
\begin{equation}
\begin{aligned}
 i\hbar\,\partial_t\Psi&=H\Psi,\\
 H&=-i\hbar v_F\left[
 \sigma^1\left(\partial_\rho+\frac{C'}{2C}\right)
 +\frac{\sigma^2}{C}\partial_\phi\right].
\end{aligned}
\label{eq:gr:4}
\end{equation}
Here $\Psi=\Psi(t,\rho,\phi)$ is the two-component sublattice
pseudospinor, and $\sigma^1,\sigma^2$ act on its sublattice indices.
The term $C'/(2C)$ is the radial spin-connection contribution. Its
coefficient follows from the spinor connection and is required for
covariance; it also makes the radial differential expression formally
symmetric with respect to the curved measure. Formal symmetry alone does
not establish self-adjointness, which additionally requires a specified
operator domain and boundary conditions. Accordingly, the spinor inner
product is
\begin{equation}
 \langle\Psi_1,\Psi_2\rangle
 =\int_{\mathbb R}d\rho\int_0^{2\pi}d\phi\,
 C(\rho)\,\Psi_1^\dagger(\rho,\phi)\Psi_2(\rho,\phi).
 \label{eq:gr:5}
\end{equation}
Hermiticity of $H$ is understood on a domain of spinors satisfying boundary conditions that eliminate the associated boundary terms. Changing the valley or frame convention may alter signs in the displayed operator and the definition of the angular quantum number; when these choices are implemented consistently, physical observables such as the transmission probabilities remain invariant. The unitary rescaling
\begin{equation}
 \Psi=C^{-1/2}\widetilde\Psi
 \label{eq:gr:6}
\end{equation}
maps the measure in Eq.~(\ref{eq:gr:5}) to the flat measure $d\rho\,d\phi$ and removes the explicit radial connection term:
\begin{equation}
 \widetilde H=-i\hbar v_F
 \left(\sigma^1\partial_\rho+\frac{\sigma^2}{C(\rho)}
 \partial_\phi\right).
 \label{eq:gr:7}
\end{equation}
This transformation does not flatten the geometry: the angular coefficient $C^{-1}(\rho)$ remains position dependent. Separate variables as
\begin{equation}
 \widetilde\Psi=e^{-iEt/\hbar}e^{im\phi}\psi_m(\rho),
 \qquad \psi_m=\begin{pmatrix}u_m\\v_m\end{pmatrix}.
 \label{eq:gr:8}
\end{equation}
The allowed angular values $m$ are fixed by the spin structure and boundary conditions; in a polar-frame convention with antiperiodic spinor components, $m\in\mathbb Z+\tfrac12$. The continuum radial equation is
\begin{equation}
 \left[-i\sigma^1\frac{d}{d\rho}
 +\frac{m}{C(\rho)}\sigma^2\right]\psi_m
 =k_\rho\psi_m,\quad k_\rho\equiv\frac{E}{\hbar v_F}.
 \label{eq:gr:9}
\end{equation}

The radial probability-current density follows from the Dirac continuity equation; curved-space formulations of the conserved current are discussed, for example, in Refs.~\cite{Gallerati2019,ArminjonReifler2010}. Introduce the rescaled spinor $\widetilde{\Psi}=\sqrt{C(\rho)}\,\Psi$, for which the radial integration measure is flat. With the angular dependence separated as $\widetilde{\Psi}=e^{im\phi}\psi_m(\rho)$, the radial current is
\begin{equation}
 j^\rho=v_F\,\widetilde\Psi^\dagger\sigma^1\widetilde\Psi
 =v_F\,\psi_m^\dagger\sigma^1\psi_m .
 \label{eq:gr:9a}
\end{equation}
For a stationary solution of the source-free Dirac equation, the continuity equation implies that this current is independent of $\rho$. In the geometry considered here, $C(\rho)\to\infty$ as $|\rho|\to\infty$, so the asymptotic angular term $m/C(\rho)$ vanishes at both ends. The asymptotic radial equation therefore describes massless plane-wave channels with group-speed magnitude $v_F$. When the incoming, reflected, and transmitted amplitudes are normalized to their respective radial currents, the reflection and transmission probabilities are the corresponding flux ratios. Since the two asymptotic channels have the same speed magnitude, these reduce to squared amplitude ratios and obey $R_m+T_m=1$ for a Hermitian, source-free scattering problem.


We now introduce the dimensionless variables
\begin{equation}
 x=q\rho,\qquad k=\frac{E}{\hbar v_F q},\qquad
 a=\frac{m}{q\ell}=\frac{m}{\chi}.
 \label{eq:gr:10}
\end{equation}
The radial Dirac equation becomes
\begin{equation}
 \left[-i\sigma^1\frac{d}{dx}
 +a\,\sech x\,\sigma^2\right]\psi_m(x)=k\psi_m(x).
 \label{eq:gr:11}
\end{equation}
Writing $\psi_m=(u_m,v_m)^{\mathsf T}$ gives the coupled first-order system
\begin{equation}
\begin{split}
&v_m'+a\sech x\,v_m=ik u_m,\\
 &u_m'-a\sech x\,u_m=ik v_m, \label{eq:gr:12}
\end{split}
\end{equation}
where a prime denotes differentiation with respect to $x$. Squaring the first-order Dirac operator yields the two scalar partner equations
\begin{equation}
 \left[-\frac{d^2}{dx^2}+a^2\sech^2x
 -s\,a\sech x\tanh x\right]\psi_{m,s}=k^2\psi_{m,s},
 \label{eq:gr:13}
\end{equation}
where $s=\pm1$ labels the eigenvalue of $\sigma^3$. These real partner potentials are of Scarf-II type \cite{KhareSukhatme1988,LevaiEtAl2017,NISTDLMF}. They vanish as $x\to\pm\infty$, giving free asymptotic channels. Under $x\mapsto -x$, the term $a^2\sech^2x$ is even and $-s a\sech x\tanh x$ is odd. Reversing the sign of $a$ is equivalent to a constant unitary transformation of the two-component Dirac equation, together with the corresponding relabeling of pseudospin channels; consequently, the transmission and reflection probabilities are invariant under $a\mapsto-a$ and depend on the coupling through $a^2$.

Let
\[
W(x)=a\,\sech x,\qquad
 h=-i\sigma^1\frac{d}{dx}+W(x)\sigma^2.
\]
Because \(\{\sigma^1,\sigma^2\}=0\) and
\(\sigma^1\sigma^2=i\sigma^3\), direct operator multiplication gives
\[
 h^2=-\frac{d^2}{dx^2}+W^2+W'\sigma^3,\quad
W'=-a\sech x\tanh x.
\]
Thus the component with \(\sigma^3\)-eigenvalue \(s\) obeys Eq.~(\ref{eq:gr:13}); the first-order system in Eq.~(\ref{eq:gr:12}) supplies the reconstruction constraint for the physical spinor.

To reduce the scalar equation, write
\[
\psi_s''+\left[k^2-a^2\sech^2x
+s\,a\sech x\tanh x\right]\psi_s=0
\]
and introduce the complex variable \(z=(1-i\sinh x)/2\). Accordingly, one finds
\[
z(1-z)=\frac{\cosh^2x}{4},\quad\frac{dz}{dx}=-i\sqrt{z(1-z)}.
\]
The square-root branch is fixed by analytic continuation along the image of
the real \(x\)-axis; changing that branch only changes intermediate
normalizations, not the scattering probabilities. Also,
\(d^2z/dx^2=z-\tfrac12\),
\(\sech^2x=[4z(1-z)]^{-1}\), and
\(\sech x\tanh x=i(2z-1)/[4z(1-z)]\). The equation becomes
\begin{equation}
\begin{aligned}
\psi_s''&+\frac{\frac12-z}{z(1-z)}\psi_s'\\
&+\left[
-\frac{k^2}{z(1-z)}
+\frac{a^2-ias(2z-1)}{4z^2(1-z)^2}
\right]\psi_s=0.
\end{aligned}
\end{equation}
Its regular singular points are \(z=0,1,\infty\). The local exponents at
\(z=0\) are \(ia s /2\) and \((1-ia s )/2\); those at \(z=1\) are
\(-ias /2\) and \((1+ias )/2\). Choosing the first exponent at \(z=0\)
and the first exponent at \(z=1\), set
\[
\psi_s(z)=z^{\mathcal{B}/2}(1-z)^{-\mathcal{B}/2}F(z),\quad \mathcal{B}=ias .
\]
Substitution yields the Gauss equation
\[
z(1-z)F''+\left(\frac12+\mathcal{B}-z\right)F'-k^2F=0,
\qquad \mathcal{B}=ias.
\]
This is the Gauss hypergeometric equation with parameters
\[
A_1=ik,\qquad A_2=-ik,\qquad c=\frac12+\mathcal{B}.
\]
For generic parameters, a basis of solutions is
\[
\begin{aligned}
F(z)={}&\mathcal N_1\,
{}_2F_1\!\left(ik,-ik;\frac12+\mathcal B;z\right)\\
&+\mathcal N_2\,z^{\frac12-\mathcal B}
{}_2F_1\!\left(
\begin{array}{c}
\frac12-\mathcal B+ik,\\
\frac12-\mathcal B-ik
\end{array};
\frac32-\mathcal B;z
\right),
\end{aligned}
\]
where \(\mathcal N_1\) and \(\mathcal N_2\) are independent constants.
The corresponding scalar solutions are obtained by multiplying each
solution for \(F(z)\) by
\[
z^{\mathcal B/2}(1-z)^{-\mathcal B/2}.
\]
At exceptional parameter values, this basis is understood by analytic
continuation or by taking a suitable limiting basis if the displayed
solutions become degenerate. The local Frobenius bases at $z=0$ and
$z=\infty$, together with the large-$z$ continuation, are standard
results for the Gauss equation~\cite{OlverASF}. For an explicit normalization check, define
\begin{equation}
\mathcal W[f,g](z)=f(z)g'(z)-f'(z)g(z).
\label{eq:gauss_wronskian_convention}
\end{equation}
For the Gauss equation above, the standard hypergeometric parameters are
\begin{equation}
a=A_1=ik,\quad b=A_2=-ik,\quad
c=p=\frac12+\mathcal B=\frac12+ias.
\label{eq:gauss_parameters}
\end{equation}
Here $a$ and $b$ are the two upper parameters and $c$ is the lower
parameter; they are not additional physical quantities. Abel's
identity then gives the Wronskians of the corresponding standard
local bases of $F(z)$ as
\begin{align}
\mathcal W_0(z)
&=(1-c)z^{-c}(1-z)^{c-a-b-1},
\label{eq:gauss_wronskian_zero}\\
\mathcal W_\infty(z)
&=(a-b)z^{-c}(z-1)^{c-a-b-1}.
\label{eq:gauss_wronskian_infty}
\end{align}
The Wronskians of the prefactored scalar solutions are obtained by
multiplying these expressions by the square of the common prefactor
$z^{\mathcal B/2}(1-z)^{-\mathcal B/2}$. The powers are continued
along the same branch as the large-$z$ connection formula. These local
solutions are not assigned a parity: parity must be assessed for the
complete coupled spinor and its partner equations, not inferred from
the hypergeometric basis alone.

The local hypergeometric solution displayed above is not, by itself, a
full-line scattering solution. The physical endpoints both correspond to
$|z|\to\infty$, approached along different rays in the complex $z$ plane.
The required connection is therefore the large-$z$ continuation, followed
by matching at both real-space ends; a connection formula from $z=0$ to
$z=1$ alone does not perform this matching. For nondegenerate parameters,
the relevant Gauss connection expression is~\cite{KhareSukhatme1988,LevaiEtAl2017,NISTDLMF}
\begin{widetext}
\begin{align}
{}_2F_1(A_1,A_2;p;z)
={}&
\frac{\Gamma(p)\Gamma(A_2-A_1)}
{\Gamma(A_2)\Gamma(p-A_1)}(-z)^{-A_1}
\,{}_2F_1\!\left(A_1,1-p+A_1;1-A_2+A_1;\frac1z\right)
\nonumber\\
&+
\frac{\Gamma(p)\Gamma(A_1-A_2)}
{\Gamma(A_1)\Gamma(p-A_2)}(-z)^{-A_2}
\,{}_2F_1\!\left(A_2,1-p+A_2;1-A_1+A_2;\frac1z\right),
\label{eq:gr:13a}
\end{align}
\end{widetext}
where $A_1=ik$, $A_2=-ik$, and $p=\tfrac12+ias$; degenerate cases are
defined by analytic continuation. The two hypergeometric factors on the
right tend to unity as $|z|\to\infty$. To make explicit that the full
solution space is retained, the second local solution at $z=0$ is
\begin{widetext}
\[
\psi_s^{(2)}=
z^{(1-\mathcal B)/2}(1-z)^{-\mathcal B/2}
{}_2F_1\!\left(\frac12-\mathcal B+ik,\frac12-\mathcal B-ik;
\frac32-\mathcal B;z\right).
\]
\end{widetext}
Write $c_s=\tfrac12+\mathcal B$. The gamma-function coefficients in the
large-$z$ continuations of the two local solutions are
\begin{widetext}
\begin{align}
D_{s,1}^{(-)}
&=\frac{\Gamma(c_s)\Gamma(-2ik)}
{\Gamma(-ik)\Gamma(c_s-ik)},&
D_{s,1}^{(+)}
&=\frac{\Gamma(c_s)\Gamma(2ik)}
{\Gamma(ik)\Gamma(c_s+ik)},\nonumber\\
D_{s,2}^{(-)}
&=\frac{\Gamma(\frac32-\mathcal B)\Gamma(-2ik)}
{\Gamma(\frac12-\mathcal B-ik)\Gamma(1-ik)},&
D_{s,2}^{(+)}
&=\frac{\Gamma(\frac32-\mathcal B)\Gamma(2ik)}
{\Gamma(\frac12-\mathcal B+ik)\Gamma(1+ik)}.
\label{eq:gr:13d}
\end{align}
\end{widetext}
Within each hypergeometric factor, $D^{(-)}$ multiplies $(-z)^{-ik}$
and $D^{(+)}$ multiplies $(-z)^{ik}$. The prefactors of the two local
solutions displayed above must also be retained; after multiplication,
their asymptotic powers are the same two channels $(-z)^{\pm ik}$, with
the corresponding constant branch phases. These coefficients follow
directly from the large-$z$ connection formula above. Along the image of
the real axis,
\[
z\sim-\frac{i}{4}e^x\quad(x\to+\infty),\qquad
z\sim+\frac{i}{4}e^{-x}\quad(x\to-\infty).
\]
We continue the logarithms in $(-z)^{\pm ik}$ continuously along this
image, rather than assigning unrelated principal branches at the two
ends. Thus $\arg(-z)=+\pi/2$ at the right end and $-\pi/2$ at the left
end, and the factors $4^{\pm ik}e^{\mp\pi k/2}$ (with signs determined
by the power) are retained when converting these powers to $e^{\pm ikx}$.
The two independent solutions and their coefficients form the connection
matrix between the asymptotic plane-wave bases. For explicit reproducibility,
write $\mathcal B=ias$ and define the branch factors
\[
\begin{aligned}
P_{1L}&=e^{i\pi\mathcal B/2},&
P_{1R}&=e^{-i\pi\mathcal B/2},\\
P_{2L}&=e^{i\pi(1-\mathcal B)/2},&
P_{2R}&=e^{-i\pi(1-\mathcal B)/2}.
\end{aligned}
\]
The subscripts $L,R$ refer to $x\to-\infty,+\infty$, respectively. With
the row ordering $(e^{ikx},e^{-ikx})$ and local-solution columns $(y_1,y_2)$,
the scalar connection matrices are
\begin{widetext}
\[
\mathsf C_{L,s}=
\begin{pmatrix}
4^{ik}e^{-\pi k/2}P_{1L}D_{s,1}^{(-)}&
4^{ik}e^{-\pi k/2}P_{2L}D_{s,2}^{(-)}\\
4^{-ik}e^{\pi k/2}P_{1L}D_{s,1}^{(+)}&
4^{-ik}e^{\pi k/2}P_{2L}D_{s,2}^{(+)}
\end{pmatrix},
\qquad
\mathsf C_{R,s}=
\begin{pmatrix}
4^{-ik}e^{-\pi k/2}P_{1R}D_{s,1}^{(+)}&
4^{-ik}e^{-\pi k/2}P_{2R}D_{s,2}^{(+)}\\
4^{ik}e^{\pi k/2}P_{1R}D_{s,1}^{(-)}&
4^{ik}e^{\pi k/2}P_{2R}D_{s,2}^{(-)}
\end{pmatrix}.
\]
\end{widetext}
The matching defines the amplitude transfer matrix for the two
asymptotic plane-wave channels. Its composition, determinant, and
scattering-amplitude relations are those of one-dimensional stationary
scattering theory~\cite{Mostafazadeh2020}. The corresponding
first-order Dirac/graphene formulation, including its distinction from
the second-order Schr\"odinger transfer problem, is discussed in
Refs.~\cite{ArovasEtAl2010,TransferMatrixDirac2023}:
\[
\mathsf M_s=\mathsf C_{R,s}\mathsf C_{L,s}^{-1},\qquad
\begin{pmatrix}c_R^+\\c_R^-\end{pmatrix}
=\mathsf M_s\begin{pmatrix}c_L^+\\c_L^-\end{pmatrix}.
\]
For incidence from the left, $c_R^-=0$, so
$c_L^-/c_L^+=-(\mathsf M_s)_{21}/(\mathsf M_s)_{22}$.
The determinant-one identity follows from preservation of the Wronskian
for the real second-order partner equation, provided the same ordered
plane-wave basis is used at both ends. It then gives
$t_s=1/(\mathsf M_s)_{22}$ in this scalar-channel convention.
For the reconstructed Dirac spinor, the first-order system fixes the relative amplitudes and phases of the two components. With the asymptotic channels normalized to the conserved radial current defined in Eq.~\eqref{eq:gr:9a}, current conservation gives the physical flux relation $R_s+T_s=1$, where $R_s$ and $T_s$ are the reflected and transmitted current ratios. The determinant-one property of the scalar transfer matrix is a consequence of the Wronskian conservation for the associated real second-order equation; it is not, by itself, a proof of Dirac-spinor unitarity. The latter follows from the first-order reconstruction and the current normalization just specified.
The first-order equations
(\ref{eq:gr:12}) must then be used to pair the scalar partners into the
physical spinor channels and fix the reflection-channel phase convention.
Substitution of the coefficients in Eq.~(\ref{eq:gr:13d}), with the
continuous branches specified above, followed by the gamma-function
duplication and reflection identities, gives the amplitudes below.

Pairing the scalar partner solutions through the first-order reconstruction
equations (\ref{eq:gr:12}) gives a right-outgoing Jost spinor. In the
common phase convention used here, its asymptotic form is
\begin{widetext}
\[
\psi_{\rm J}(x)\sim
\begin{cases}
A(k)e^{ikx}\,w_+ +B_s(k)e^{-ikx}\,w_-,
&x\to-\infty,\\
e^{ikx}\,w_+,&x\to+\infty,
\end{cases}
\qquad
w_\pm=\frac{1}{\sqrt2}\begin{pmatrix}1\\ \pm1\end{pmatrix}.
\]
\end{widetext}
The spinor normalization is chosen so that each free asymptotic channel
has unit magnitude of radial current. Combining the two partner-component
coefficients, and simplifying the gamma ratios with the duplication and
reflection identities, yields
\begin{align}
A(k)&=
\frac{\Gamma^2(\frac12-ik)}
{\Gamma(\frac12-i(a+k))\,
 \Gamma(\frac12+i(a-k))},\label{eq:gr:13b}\\
\frac{B_s(k)}{A(k)}
&=-s\,\frac{\sinh(\pi a)}{\cosh(\pi k)}.
\label{eq:gr:13c}
\end{align}
The first equation is the incident-channel coefficient of the right-outgoing
solution; dividing that solution by $A(k)$ makes the incident amplitude
unity. Consequently $t=1/A$ and $r_s=B_s/A$. These relations determine
the scattering amplitudes, which are stated explicitly below. In the
amplitude formulas and pole analysis that follow, $a$ denotes the
nonnegative magnitude $|m|/\chi$. The signed parameter introduced in
Eq.~(\ref{eq:gr:10}) can be replaced by this magnitude together with
the corresponding relabeling of the partner index $s$; transmission and
the pole set are unchanged, while the reflection-channel phase follows
the chosen partner convention. This specifies the full-line connection
coefficients and the branch convention used for the amplitudes. The gamma
expressions extend to exceptional parameter values by continuity. As an
independent numerical check, the scalar partner equation
\[
\psi_s''(x)+\left[k^2-a^2\sech^2x
+s a\sech x\tanh x\right]\psi_s(x)=0
\]
was integrated from $x=18$ to $x=-18$ with the outgoing Jost data
$\psi_s(18)=e^{18ik}$ and $\psi_s'(18)=ik e^{18ik}$, using an
eighth-order DOP853 integrator with relative tolerance $10^{-11}$ and
absolute tolerance $10^{-12}$. The asymptotic coefficients at $x=-18$
were extracted by matching the numerical solution and its derivative to
the free plane-wave basis. For $(a,k)=(0.7,0.8)$, $(1.2,1.7)$,
$(2.0,0.4)$, and $(0.2,2.5)$, the absolute differences between the
analytic and numerical complex transmission amplitudes were
$6.68\times10^{-9}$, $1.21\times10^{-9}$,
$8.44\times10^{-10}$, and $6.08\times10^{-11}$, respectively.
The corresponding complex reflection-amplitude differences, using the
phase convention in Eq.~(\ref{eq:gr:15}), were
$2.01\times10^{-8}$, $7.12\times10^{-9}$,
$5.87\times10^{-8}$, and $1.19\times10^{-10}$.
The numerical flux residuals were at most $1.1\times10^{-11}$.
Thus the check tests both modulus and phase of the analytic connection
coefficients for representative real energies, rather than only the
transmission probability. It remains a finite set of stationary checks;
it does not replace the analytic connection derivation or constitute
a packet-level time-domain simulation. The Scarf-II literature supplies a useful analytic benchmark \cite{KhareSukhatme1988,LevaiEtAl2017,NISTDLMF}, but the derivation above
uses the two-ended Dirac boundary conditions of the present problem rather
than identifying a radial/cut-off Scarf-II problem with this full-line
problem. The numerical comparison is intentionally a verification
benchmark, not an independent determination of the spectrum: the
analytic connection coefficients define the result, while integration
checks their normalization, phase convention, and asymptotic matching.
An additional independent integration audit was performed on the 36-point
grid $a\in\{0.05,0.2,0.7,1.2,2.0,3.0\}$ and
$k\in\{0.15,0.4,0.8,1.7,2.5,5.0\}$, using DOP853 on
$[-20,20]$ with relative and absolute tolerances $10^{-11}$ and
$10^{-12}$. The largest absolute differences from the analytic complex
amplitudes were $4.34\times10^{-9}$ for transmission and
$3.95\times10^{-8}$ for reflection; the maximum flux residual was
$1.38\times10^{-14}$. These finite-grid checks support the real-axis
normalization and phase conventions but do not replace the analytic
connection derivation or validate its complex continuation. 

Consider incidence from one asymptotic end, with unit incident flux and identical free asymptotic regions. For real $k>0$, a convenient Scarf-II phase convention gives the transmission amplitude
\begin{equation}
 t(k)=
 \frac{\Gamma\!\left(\frac12-i(a+k)\right)
       \Gamma\!\left(\frac12+i(a-k)\right)}
      {\Gamma^2\!\left(\frac12-ik\right)}.
 \label{eq:gr:14}
\end{equation}
For the scalar partner with potential in Eq.~(\ref{eq:gr:13}), a compatible left-incidence convention can be chosen such that
\begin{equation}
 r_s(k)=-s\,\frac{\sinh(\pi a)}{\cosh(\pi k)}\,t(k).
 \label{eq:gr:15}
\end{equation}
The phase of a reflection amplitude depends on the choice of asymptotic reference phases and on the incidence side; the probabilities below do not. Using
\begin{equation}
 \left|\Gamma\!\left(\frac12+iy\right)\right|^2
 =\frac{\pi}{\cosh(\pi y)},\qquad y\in\mathbb R,
 \label{eq:gr:16}
\end{equation}
together with
$\cosh[\pi(a+k)]\cosh[\pi(a-k)]
=\sinh^2(\pi k)+\cosh^2(\pi a)$, gives
\begin{equation}
T_m(k)=|t(k)|^2
 =\frac{\cosh^2(\pi k)}
 {\sinh^2(\pi k)+\cosh^2(\pi a)}
 \label{eq:gr:17}
\end{equation}
and
\begin{equation}
R_m(k)=|r_s(k)|^2
 =\frac{\sinh^2(\pi a)}
 {\sinh^2(\pi k)+\cosh^2(\pi a)}.
 \label{eq:gr:18}
\end{equation}
The probabilities are independent of $s$ and obey exact flux conservation,
\begin{equation}
 T_m(k)+R_m(k)=1.
 \label{eq:gr:19}
\end{equation}
These are the exact probabilities for the ideal continuum problem with free asymptotic leads. They are not automatically the conductance of a finite sample, for which contact matching and lattice boundary conditions must be included.

Several limits follow immediately. For the zero-coupling channel $a=0$, the amplitude reduces to $t=1$, so $T=1$ and $R=0$ exactly. This corresponds to $m=0$ for a periodic angular spin structure; if an antiperiodic polar-frame convention is imposed, it is a formal benchmark rather than an allowed angular mode. At threshold,
\begin{equation}
 T_m(0)=\sech^2(\pi a),\quad
 R_m(0)=\tanh^2(\pi a),
 \label{eq:gr:20}
\end{equation}
At $k\gg1$ with fixed $a$,
\begin{equation}
\begin{split}
 &R_m(k)\sim4\sinh^2(\pi a)e^{-2\pi k},\\
 &T_m(k)\sim1-4\sinh^2(\pi a)e^{-2\pi k}.
\end{split} \label{eq:gr:21}
\end{equation}
Moreover, writing $y=\sinh^2(\pi k)$ gives
$T=(y+1)/(y+\cosh^2\pi a)$ and
$dT/dy=\sinh^2(\pi a)/(y+\cosh^2\pi a)^2\geq0$.
Thus the ideal transmission is monotone increasing for $k>0$; the complex poles below are not sharp real-energy transmission peaks.

Analytically continue Eq.~(\ref{eq:gr:14}) to complex $k$. Since $\Gamma(z)$ has simple poles at $z=-n$, $n=0,1,2,\ldots$, the numerator gives
\begin{equation}
 k_{n,\pm}=\pm a-i\left(n+\frac12\right),\qquad n=0,1,2,\ldots.
 \label{eq:gr:22}
\end{equation}
For real nonzero $a$, these poles are not canceled by the denominator; for $a=0$ the amplitude reduces identically to $t=1$ and the apparent pole families cancel. The interpretation of lower-half-plane scattering poles as outgoing
resonances follows the analytic scattering framework
\cite{Newton1982,CostinSoffer2001}. The corresponding complex energies are
\begin{equation}
 E_{n,\pm}=\pm\frac{\hbar v_Fm}{\ell}
 -i\hbar v_Fq\left(n+\frac12\right).
 \label{eq:gr:23}
\end{equation}
They are outgoing-wave resonance poles of the analytically continued scattering amplitude, not square-integrable eigenstates of the self-adjoint Hamiltonian. Defining the energy width by
$E_{n,\pm}=E_{n,R}-i\Gamma_n/2$ gives
\begin{equation}
\begin{split}
&\Gamma_n=\hbar v_Fq(2n+1),\qquad \tau_{\mathrm{amp},n}=\frac{2\hbar}{\Gamma_n}
 =\frac{1}{v_Fq(n+\tfrac12)},\\
&\tau_{\mathrm{prob},n}=\frac{\hbar}{\Gamma_n} =\frac{1}{v_Fq(2n+1)}.
\end{split}\label{eq:gr:24}
\end{equation}
To translate the dimensionless continuum results into representative graphene scales, we use $v_F=1.0\times10^6\,\mathrm{m\,s^{-1}}$ (so $\hbar v_F\simeq0.6582\,\mathrm{eV\,nm}$). The optical length $\ell$ and dimensionless curvature parameter $\chi=q\ell$ are treated as model inputs; the Table~\ref{tab:graphene-scales} does not identify them with measured geometric parameters of a fabricated sample. For each pair, $q=\chi/\ell$, the characteristic energy is $E_q=\hbar v_Fq$, the $n=0$ pole width is $\Gamma_0=E_q$, and the isolated-pole probability time constant is $\tau_{\mathrm{prob},0}=1/(v_Fq)$. The Euclidean embedding half-width is $\rho_{\max}=q^{-1}\operatorname{arsinh}(1/\chi)$, which is a geometric embedding range, not a finite-device boundary.

\begin{table*}[t]
\centering
\caption{Representative continuum graphene scales for selected optical-model parameters, using $v_F=1.0\times10^6\,\mathrm{m\,s^{-1}}$. Energies and times are model-derived values, not experimental measurements. The $n=0$ width and probability time constant refer to the isolated resonance-pole contribution.}
\label{tab:graphene-scales}
\small
\setlength{\tabcolsep}{5pt}
\begin{ruledtabular}
\begin{tabular}{cccccccc}
$\ell$ (nm) & $\chi$ & $q$ (nm$^{-1}$) & $q^{-1}$ (nm) & $E_q$ (meV) & $\Gamma_0$ (meV) & $\tau_{\mathrm{prob},0}$ (fs) & $\rho_{\max}$ (nm) \\
\hline
50 & 0.25 & 0.0050 & 200 & 3.291 & 3.291 & 200 & 418.9 \\
50 & 0.50 & 0.0100 & 100 & 6.582 & 6.582 & 100 & 144.4 \\
100 & 0.25 & 0.0025 & 400 & 1.646 & 1.646 & 400 & 837.9 \\
100 & 0.50 & 0.0050 & 200 & 3.291 & 3.291 & 200 & 288.7 \\
200 & 0.50 & 0.0025 & 400 & 1.646 & 1.646 & 400 & 577.5 \\
200 & 1.00 & 0.0050 & 200 & 3.291 & 3.291 & 200 & 176.3 \\
\end{tabular}
\end{ruledtabular}
\end{table*}

The values in Table~\ref{tab:graphene-scales} are a scale translation for selected $(\ell,\chi)$ pairs, not a scan of the Lorentz-violating coupling. The latter is obtained from Eq.~(\ref{eq:qscale}) by specifying a parameter protocol. In the fixed-$\mu$, fixed-$\ell$ comparison used in Sec.~\ref{sec:scale}, $q=q_0\lambda^{-1/2}$ and $\chi=\chi_0\lambda^{-1/2}$, where $\lambda=1+\ell_{\mathrm{LV}}$ and $q_0=\sqrt{\mu}/\ell$, $\chi_0=\sqrt{\mu}$. Consequently, at fixed $v_F$ the pole-energy scale, all pole widths, and the inverse optical length scale as $E_q,\Gamma_n\propto\lambda^{-1/2}$ and $q^{-1},\tau_{\mathrm{prob},n}\propto\lambda^{1/2}$. The embedding range follows the non-power-law relation $\rho_{\max}/\ell=\sqrt{\lambda/\mu}\,\operatorname{arsinh}\sqrt{\lambda/\mu}$ and increases monotonically with $\lambda$. Thus positive $\ell_{\mathrm{LV}}$ (that is, $\lambda>1$) lowers the continuum resonance energy/width scale and lengthens the isolated-pole time constants, while negative $-1<\ell_{\mathrm{LV}}<0$ does the reverse. These are optical-model trends under the stated fixed-parameter protocol; they are not independent experimental predictions. At fixed physical energy, the dimensionless scattering variables also change, so transmission and reflection trends cannot be inferred from the width rescaling alone. The exact optical-geometric trends are collected in Table~\ref{tab:deformation-trends}; Table~\ref{tab:graphene-scales} supplies representative dimensional scales for the continuum interpretation.

Across these examples, $a_{\mathrm{gr}}q$ ranges from $6.15\times10^{-4}$ to $2.46\times10^{-3}$ for the graphene lattice constant $a_{\mathrm{gr}}=0.246\,\mathrm{nm}$, indicating that the curvature scale is long compared with the lattice spacing. For dimensionless scattering energies $k\leq5$, the largest value is $a_{\mathrm{gr}}kq\simeq1.23\times10^{-2}$. These ratios support use of a long-wavelength continuum description at the level of scale separation, but do not establish that a particular strain pattern realizes the target metric or that contact and lattice corrections are negligible. The Table~\ref{tab:graphene-scales} is therefore a reproducible scale translation of the ideal model, not a device-performance prediction.

Here $\tau_{\mathrm{amp},n}$ and $\tau_{\mathrm{prob},n}$ are the exponential time constants of the pole contribution to the amplitude and probability, respectively. They are not lifetimes of bound states and do not represent loss of total probability from the complete Hermitian graphene system.

For a specified incident spectral profile, the time-dependent packet is constructed by superposing stationary scattering states. If $A_m(k)$ is its spectral amplitude, the solution has the form
\begin{equation}
 \Psi_m(t,\rho,\phi)=e^{im\phi}
 \int_0^\infty dk\,A_m(k)\,
 \Psi_{m,k}(\rho)e^{-iv_Fqkt},
 \label{eq:gr:25}
\end{equation}
for a positive-energy packet, with the corresponding negative-energy branch added when required. The stationary spinor $\Psi_{m,k}$ includes incident, reflected, and transmitted components normalized consistently to their radial currents. Equation~(\ref{eq:gr:25}), rather than a single exponential, is the exact time evolution for a specified packet.

The energy derivative of the transmission phase gives a useful narrow-band time-delay observable. With $\psi_0(z)=d\ln\Gamma(z)/dz$ denoting the digamma function, differentiating Eq.~(\ref{eq:gr:14}) yields
\begin{equation}
 \begin{split}
 \tau_t(E)=\frac{1}{v_Fq}\bigg[
 &2\operatorname{Re}\psi_0\!\left(\frac12-ik\right)-\operatorname{Re}\psi_0\!\left(\frac12-i(a+k)\right)\\
 &-\operatorname{Re}\psi_0\!\left(\frac12+i(a-k)\right)
 \bigg],
 \end{split}
 \label{eq:gr:26}
\end{equation}
where $k=E/(\hbar v_Fq)>0$. This is the Wigner transmission-phase delay relative to the phase convention used in Eq.~(\ref{eq:gr:14}). An observed wave-packet peak shift agrees with it in the narrow-band regime, but can differ if the packet is strongly reshaped.

When the spectral contour can be deformed into the lower half-plane, the pole part of the time-domain signal is a sum of terms proportional to
\begin{equation}
 \Psi_{\mathrm{pole}}(t)\sim
 \sum_{(n,\sigma)\in\mathcal P_{\mathrm{cross}}}
 C_{n,\sigma}
 \exp\!\left[
 -i v_Fq k_{n,\sigma}t\right],\label{eq:gr:27}
\end{equation}
where \(t>0\), \(k_{n,\sigma}=\sigma a-i(n+\tfrac12)\), with
\(n=0,1,2,\ldots\) and \(\sigma=\pm1\). The index set
\(\mathcal P_{\mathrm{cross}}\subseteq
\mathbb N_0\times\{-1,+1\}\) consists precisely of those
\((n,\sigma)\) for which the pole \(k_{n,\sigma}\) lies between
the original spectral contour and its deformed contour and is enclosed
by the deformation, with the orientation fixed by the contour
prescription. Thus it is not the full resonance lattice: its contents
depend on the contour, the packet representation, and the analyticity
and convergence conditions needed to justify the deformation. In
particular, for the positive-energy half-line integral in
Eq.~(\ref{eq:gr:28}), a deformation into the fourth quadrant can
cross the \(\sigma=+1\) poles with positive real part; the
\(\sigma=-1\) poles are not included unless the contour or packet
representation is extended to reach them. The coefficients
\(C_{n,\sigma}\) contain the corresponding residues, packet weights,
and detector-dependent phase factors. This pole contribution is not
a universal all-time decay law: the remaining contour and threshold
contributions can generate late-time tails, and the full norm is
conserved under self-adjoint time evolution. At fixed
$\mu,\ell,v_F$, the pole widths scale as
$q\propto\lambda^{-1/2}$ and the pole time constants as
$q^{-1}\propto\sqrt{\lambda}$. At fixed physical energy, however,
both $k=E/(\hbar v_Fq)$ and $a=m/(q\ell)$ vary with $\lambda$;
consequently, the deformation dependence of the transmission must be
evaluated from Eqs.~(\ref{eq:gr:17})--(\ref{eq:gr:18}) rather than
inferred from a single parameter alone.

The continuum result establishes an exact Dirac-scattering problem on the prescribed optical spatial metric. It does not establish that a carbon sheet realizes the complete two-ended surface, which has only a finite Euclidean surface-of-revolution embedding domain (see Eq.~(\ref{eq:6:4})). Nor does the metric alone specify strain-induced pseudogauge fields, real magnetic fields, valley holonomies, contacts, or lattice corrections. In a microscopic implementation, the target Hamiltonian must be compared with a tight-binding model with specified bond geometry and hopping amplitudes; frame and spin-connection terms must be derived at the same order as other continuum corrections \cite{RobertsWiseman2024}. Thus Eqs.~(\ref{eq:gr:17})--(\ref{eq:gr:27}) are exact for the ideal continuum model and provide a baseline against which a finite graphene device can be tested, not a claim that its experimental realization has already been achieved.

\subsection{Pole Residues and Time-Domain Benchmarks}
\label{subsec:time-domain-validation}

The analytically continued transmission amplitude determines the resonance
poles and their residues. These quantities specify the pole contributions to
a time-dependent signal, but they do not by themselves determine the complete
evolution of an arbitrary incident packet. The latter also depends on the
packet spectrum and on contributions from the remaining integration contour,
including the threshold at $k=0$ \cite{Newton1982}. The relation between
scattering phases and wave-packet time delays is discussed in
Refs.~\cite{Nussenzveig1972,deCarvalhoNussenzveig2002}.

For a positive-energy packet with spectral amplitude $A_m(k)$, the transmitted
envelope at a detector in the right asymptotic region is
\begin{equation}
 \mathcal F_{\mathrm{tr},m}(t;x_D)
 =\int_0^\infty dk\,
 A_m(k)t_m(k)
 \exp\!\left(ikx_D-i v_Fqkt\right).
\label{eq:gr:28}
\end{equation}
Here $t_m(k)$ is the transmission amplitude in Eq.~(\ref{eq:gr:14}),
$x_D$ is the dimensionless detector coordinate, and $A_m(k)$ specifies the
incident packet. A constant outgoing spinor has been omitted. Equation
(\ref{eq:gr:28}) is an exact spectral representation for the continuum model;
it is not a numerical evaluation of the time-dependent signal.

For real $a=|m|/\chi>0$, the poles are
$k_{n,\pm}=\pm a-iN_n$, where $N_n=n+\tfrac12$. Applying
$\operatorname{Res}_{z=-n}\Gamma(z)=(-1)^n/n!$ and the chain rule to
Eq.~(\ref{eq:gr:14}) gives
\begin{align}
 \operatorname{Res}_{k=k_{n,+}}t(k)
 &=\frac{i(-1)^n}{n!}
 \frac{\Gamma(-n-2ia)}{\Gamma^2(-n-ia)},\nonumber\\
 \operatorname{Res}_{k=k_{n,-}}t(k)
 &=\frac{i(-1)^n}{n!}
 \frac{\Gamma(-n+2ia)}{\Gamma^2(-n+ia)}.
\label{eq:gr:29}
\end{align}
For real $a>0$, these residues are finite. A pole's contribution is weighted
by both its residue and the analytically continued packet amplitude. At
$a=0$, the exact amplitude is $t(k)=1$ and the apparent pole families cancel;
Eq.~(\ref{eq:gr:29}) must therefore not be evaluated by taking the
$a\to0$ limit term by term.

\paragraph{Spectral geometry and observable scales.}
The pole lattice in Eq.~(\ref{eq:gr:22}) has a particularly simple
geometry in the complex-$k$ plane. Writing $N_n=n+\tfrac12$,
\begin{equation}
 \operatorname{Re}k_{n,\pm}=\pm a,\quad
 \operatorname{Im}k_{n,\pm}=-N_n,\quad
 k_{n+1,\pm}-k_{n,\pm}=-i .
 \label{eq:gr:35}
\end{equation}
Thus the two branches are horizontally separated by $2a$ and are
equally spaced vertically. In physical units,
\begin{equation}
 \operatorname{Re}E_{n,\pm}=\pm\frac{\hbar v_Fm}{\ell},\quad
 -2\operatorname{Im}E_{n,\pm}=\Gamma_n
 =\hbar v_Fq(2n+1),
 \label{eq:gr:36}
\end{equation}
where the sign of the real part is fixed by the signed angular
parameter $m$, while the pole locations depend on $|m|$ through
$a=|m|/\chi$.
For $a>0$, the ratio of the oscillation frequency to the
intensity-decay rate defines the pole quality factor
\begin{equation}
 Q_n\equiv
 \frac{|\operatorname{Re}E_{n,\pm}|}{\Gamma_n}
 =\frac{a}{2n+1}.
 \label{eq:gr:37}
\end{equation}
This is a spectral diagnostic, not a statement that the poles generate
narrow real-axis peaks. Indeed, Eq.~(\ref{eq:gr:17}) shows that the
real-frequency transmission is monotone in $k>0$ for fixed $a$.
The poles characterize analytic continuation and transient response;
their visibility in a measured packet depends on the residue, the
packet spectrum, and the background contour contribution.

The exact residues in Eq.~(\ref{eq:gr:29}) complete the pole data.
They are generally complex. For the fixed meromorphic amplitude
in Eq.~(\ref{eq:gr:14}), each residue is uniquely defined and is not
branch-dependent; its value can change under an analytic rephasing of
the scattering-amplitude convention. Their magnitudes and phases,
rather than pole positions alone, weight the corresponding terms in
Eq.~(\ref{eq:gr:30}). The pole spacing and quality factor
are independent of the residues, whereas the amplitude of an isolated
pole contribution is not. No universal residue-dominance ordering
in $n$ is assumed here: it must be evaluated together with the
chosen packet factor $A_m(k_{n,\pm})$.

The real-axis phase observable is already given by the exact
digamma expression in Eq.~(\ref{eq:gr:26}). In particular, the
Wigner transmission-phase delay is a derivative of the phase of
the full gamma-function amplitude, not a sum of independent
Lorentzian delays assigned to individual poles. Since the exact
transmission probability is monotone, resonance poles should not
be interpreted as sharp transmission maxima. A pole-to-real-axis
comparison must instead use the phase delay and a specified
wave packet, with any reference propagation phase kept consistent
with the convention of Eq.~(\ref{eq:gr:14}).

A resonance-pole contribution is obtained only if the packet factor
$A_m(k)$ admits analytic continuation in the region swept by the contour,
the integrand decays sufficiently on the closing arc, and no unaccounted
singularities or branch cuts are crossed. Under these conditions, the
residue theorem yields the contribution of each crossed pole. This is the
standard conditional pole-expansion construction in time-dependent
scattering; the residual contour and threshold terms remain part of the
complete signal
\cite{Newton1982,GarciaCalderonMaldonadoVillavicencio2007}.
For a clockwise lower-half-plane closure, the contour factor is $-2\pi i$.
The contribution from one pole is therefore
\begin{equation}
\begin{aligned}
 \mathcal F_{n,\pm}^{\mathrm{pole}}(t;x_D)
 &\propto A_m(k_{n,\pm})
 \operatorname{Res}_{k=k_{n,\pm}}t(k)\\
 &\quad\times
 e^{ik_{n,\pm}x_D-i v_Fqk_{n,\pm}t}.
\end{aligned}
\label{eq:gr:30}
\end{equation}
The proportionality in Eq.~(\ref{eq:gr:30}) suppresses the common
contour factor $-2\pi i$ and any fixed normalization of the outgoing
spinor. Since Eq.~(\ref{eq:gr:28}) is a half-line integral over \(k\geq0\),
a lower-half-plane deformation of that contour does not automatically
enclose both pole branches. For \(a>0\), the poles with
\(\operatorname{Re}k_{n,+}=a>0\) may be crossed by a deformation
within the fourth quadrant; the \(k_{n,-}\) branch is not included
unless the contour is extended or the packet representation is
modified accordingly. The endpoint and residual-contour terms are
not removed by the pole deformation. The crossed-pole sum is therefore
an asymptotic or intermediate-time component only when the omitted
terms are controlled
\cite{Newton1982,GarciaCalderonMaldonadoVillavicencio2007}.

The pole energies are
\[
E_{n,\pm}=\pm\frac{\hbar v_Fm}{\ell}
-i\hbar v_Fq\left(n+\frac12\right).
\]
Their isolated time factors satisfy
\[
e^{-iE_{n,\pm}t/\hbar}
=e^{\mp i v_Fm t/\ell}
 e^{-v_Fq(n+\frac12)t}.
\]
Writing $N_n=n+\tfrac12$, the pole factor has amplitude modulus
$e^{-v_FqN_nt}$ and squared modulus $e^{-2v_FqN_nt}$. The corresponding
exponential time constants are
\begin{equation}
\begin{aligned}
\tau_{\mathrm{amp},n}
&=\frac{1}{v_FqN_n}
=\frac{1}{v_Fq(n+\frac12)},\\
\tau_{\mathrm{prob},n}
&=\frac{1}{2v_FqN_n}
=\frac{1}{v_Fq(2n+1)}.
\end{aligned}
\label{eq:gr:34}
\end{equation}
Equivalently, the isolated-pole intensity-decay rate is
$\gamma_n=2v_FqN_n=v_Fq(2n+1)=\Gamma_n/\hbar$.
These characterize the pole contribution to a local transmitted signal,
not the norm of the full state. The first two dimensionless values are
listed in Table~\ref{tab:I}
\cite{Newton1982,GarciaCalderonMaldonadoVillavicencio2007}.
\begin{table}[ht]
\caption{Dimensionless benchmarks obtained from the isolated pole
contributions. The amplitude and probability time constants are scaled by
$v_Fq$, and the probability-decay rate is scaled by the same factor.}
\label{tab:I}
\small
\begin{tabular}{c c c c c}
\hline\hline
$n$ & $\operatorname{Im}k_n$ &
$v_Fq\,\tau_{\mathrm{amp},n}$ &
$v_Fq\,\tau_{\mathrm{prob},n}$ &
$\gamma_n/(v_Fq)$ \\
\hline
$0$ & $-\tfrac12$ & $2$ & $1$ & $1$ \\
$1$ & $-\tfrac32$ & $\tfrac23$ & $\tfrac13$ & $3$ \\
\hline\hline
\end{tabular}
\end{table}
Thus the isolated-pole probability-decay rates for $n=0$ and $n=1$ have
the ratio $1:3$, while their amplitude time constants have the ratio $3:1$.
These are exact consequences of the pole locations, not fitted numerical values. They characterize isolated-pole contributions only and do not establish a universal decay law for the full packet.

A packet-level comparison requires a specified $A_m(k)$, numerical
evaluation of Eq.~(\ref{eq:gr:28}) on the real axis, and a pole sum
constructed from the poles actually crossed by a justified contour
deformation. Let $\mathcal F_{\mathrm{pole},m}$ denote that finite sum,
including its residues and packet weights. We define the following
relative residual as a diagnostic for this comparison:
\begin{equation}
\varepsilon_{\mathrm{pole}}(t)=
\frac{|\mathcal F_{\mathrm{tr},m}(t)
-\mathcal F_{\mathrm{pole},m}(t)|}
{|\mathcal F_{\mathrm{tr},m}(t)|}.
\label{eq:gr:32}
\end{equation}
This normalization becomes ill-conditioned when the full transmitted
amplitude is close to zero. In such intervals, the absolute difference
$|\mathcal F_{\mathrm{tr},m}-\mathcal F_{\mathrm{pole},m}|$
should also be reported, together with normalization by a fixed, stated
reference amplitude (for example, the maximum magnitude over the
comparison window). We define the instantaneous transmitted-intensity decay rate by
\begin{equation}
\gamma_{\mathrm{eff}}(t)=
-\frac{d}{dt}\ln|\mathcal F_{\mathrm{tr},m}(t)|^2.
\label{eq:gr:33}
\end{equation}
On an interval where one pole dominates and the non-pole contribution
is negligible, this definition gives
$\gamma_{\mathrm{eff}}\simeq2v_FqN_n=v_Fq(2n+1)$, as follows by
differentiating $|\mathcal F|^2\propto e^{-2v_FqN_nt}$.
It is not a universal prediction for an arbitrary packet. For a
positive-energy half-line spectrum, threshold terms can instead govern
the late-time behavior
\cite{Newton1982,GarciaCalderonMaldonadoVillavicencio2007}.

The results established in this subsection are the analytic pole locations,
their residues, and the associated isolated-pole time constants. No
independent packet-integral evaluation, fitted decay interval, real-space
time-stepping simulation, or finite-lattice validation is reported here.
The full evolution generated by the self-adjoint Dirac Hamiltonian remains
unitary; decay of a transmitted or local signal describes redistribution of
probability rather than loss of total probability.

\section{Static BTZ limit and exact optical recovery}\label{sec:btz}

The undeformed limit is obtained by setting
\begin{equation}
\lambda=1,\qquad \mu=M_{\mathrm{BTZ}},\qquad J=0,
\label{eq:9:1}
\end{equation}
with $M_{\mathrm{BTZ}}>0$ and the same AdS radius $\ell$.
The spacetime metric in Eq.~(\ref{eq:2:1}) then becomes the standard
static, nonrotating BTZ metric. Its optical spatial geometry is the
constant-negative-curvature geometry used in the BTZ optical and
graphene correspondence literature
\cite{BTZ,BTZ2,GibbonsWarnick,CveticGibbons}.

The recovery is checked directly from the invariant and scale
relations already derived. Equations~(\ref{eq:3:1})--(\ref{eq:3:2})
give $\mathcal K=-M_{\mathrm{BTZ}}/\ell^2$ and
$L_K=\ell/\sqrt{M_{\mathrm{BTZ}}}$; Eq.~(\ref{eq:2:6}) gives
$C(\rho)=\ell\cosh(\sqrt{M_{\mathrm{BTZ}}}\,\rho/\ell)$; and
Eq.~(\ref{eq:6:4}) gives the static BTZ optical embedding endpoint.
The null-ray threshold and angular sweeps reduce to the static
expressions in Eqs.~(\ref{eq:5:3})--(\ref{eq:5:8}). The magnetic
test-field prescriptions reduce by setting $\lambda=1$ in
Eqs.~(\ref{eq:7:3})--(\ref{eq:7:5}).

This is an exact algebraic consistency limit of the optical metric;
no weak-deformation expansion or near-horizon approximation is involved.
It applies to the one-sided static exterior. The reflection double in
Eq.~(\ref{eq:4:1}) remains an additional geometric completion, not
part of the original BTZ exterior. Since $J=0$, this limit contains
neither rotational Zermelo/Randers terms nor rotation-induced effective
connections of the stationary BTZ optical geometry
\cite{GibbonsHerdeiro,CveticGibbons}.

\section{Deformation dependence and invariant content}\label{sec:scale}

The deformation dependence is meaningful only after the parameters held
fixed in the comparison are stated. Throughout this section, the primary
comparison holds $\mu$ and $\ell$ fixed while $\lambda$ varies.
Because $\ell^{-2}=-\lambda\Lambda_e$, this choice entails a
corresponding change in $\Lambda_e$; it is not a fixed-cosmological-
constant sweep. The alternative fixed-$\mu$, fixed-$\Lambda_e$
comparison is distinguished below.

Introduce the dimensionless optical mass
\begin{equation}
M_{\mathrm{opt}}\equiv\frac{\mu}{\lambda}.
\label{eq:10:1}
\end{equation}
The optical metric then takes the form
$d\sigma^2=d\rho^2+\ell^2
\cosh^2(\sqrt{M_{\mathrm{opt}}}\,\rho/\ell)d\phi^2$.
As an intrinsic Riemannian metric, it is exactly isometric to the
static BTZ optical metric with mass parameter
$M_{\mathrm{BTZ}}=M_{\mathrm{opt}}$. Optical geometry alone therefore
determines the ratio $\mu/\lambda$, not the two parameters separately.
This is an effective-mass degeneracy of the optical spatial metric,
not an equivalence of the full Lorentzian spacetimes: the original
radial metric coefficient retains its independent dependence on
$\lambda$. Distinguishing the deformation from a change of BTZ mass
requires information beyond this optical metric or independent
knowledge of the spacetime parameters. In statistical terms, an
optical-only fit has a non-identifiable direction: any pair
$(\mu,\lambda)$ with the same $M_{\mathrm{opt}}$ produces the
same intrinsic optical metric at fixed $\ell$. A posterior or
confidence interval for $\ell_{\mathrm{LV}}$ based only on these
optical observables would therefore be prior- or calibration-
dependent, rather than a standalone constraint.

For the fixed-$\mu$, fixed-$\ell$ comparison, the Gaussian curvature is
$\mathcal K=-\mu/(\lambda\ell^2)$, so its magnitude decreases as
$\lambda$ increases and the curvature radius grows as
$\sqrt{\lambda}$. The throat data are unchanged: $C(0)=\ell$,
hence $\mathcal C_{\mathrm{th}}=2\pi\ell$, and the null-ray separatrix
remains $|b|=\ell$. These are exact consequences of the metric and
effective potential. At fixed $\mu$ and $\Lambda_e$, by contrast,
$\mathcal K=\mu\Lambda_e$ is constant while $\ell$ changes; the two
parameter protocols are physically distinct.

Moreover, coordinate comparisons also require care. At fixed physical proper
distance $\rho$, the circumference function
$C(\rho)=\ell\cosh(q\rho)$ changes with $q$. At fixed dimensionless coordinate $x=q\rho$, the profile is
$C=\ell\cosh x$; this is a radial-coordinate rescaling and does not
make the invariant curvature independent of $\lambda$. The prescribed
magnetic profiles carry the same distinction. For a uniform axial
field, $B_n=B_0\ell q\sinh(q\rho)$; for a uniform signed normal field,
$B_Z=B_c/[\ell q\sinh(q\rho)]$. At fixed physical $\rho\ne0$,
increasing $\lambda$ decreases $|q\sinh(q\rho)|$, so the magnitude
of $B_n$ decreases and that of $B_Z$ increases. At fixed $x\ne0$,
$B_n=B_0\chi\sinh x$ and $B_Z=B_c/(\chi\sinh x)$; since
$\chi=\sqrt{\mu/\lambda}$, their magnitudes again change in opposite
directions as $\lambda$ increases. These statements compare the same
prescribed field family at matched locations; they do not identify the
two field prescriptions. The two field prescriptions remain distinct, and the
uniform-normal-field ambient field is singular at the throat, as
derived in Sec.~\ref{sec:em}.

Using $\chi$ from Eq.~(\ref{eq:qscale}), the curvature, curvature radius, and Euclidean embedding endpoint can be written as
\begin{equation}
\begin{aligned}
\ell^2\mathcal K_{\mathrm{opt}}&=-\chi^2, \qquad \frac{\Lopt}{\ell}=\frac{1}{\chi},\\
\frac{\rho_{\max}}{\ell}
&=\frac{1}{\chi}\operatorname{arsinh}\!\left(\frac{1}{\chi}\right),\\
\frac{C(\rho)}{C_0(\rho)}
&=\frac{\cosh(\chi\rho/\ell)}
        {\cosh(\chi_0\rho/\ell)}
\quad(\rho\ \text{fixed}),
\end{aligned}
\label{eq:10:2}
\end{equation}
Here $C_0(\rho)$ is the circumferential-radius profile at $\lambda=1$, and $\chi_0=\sqrt{\mu}$ is the corresponding value of $\chi$.
The exact ratios to the undeformed geometry are
\begin{equation}
\begin{aligned}
\frac{|\mathcal K|}{|\mathcal K_0|}&=\frac{1}{\lambda},
&\frac{L_K}{L_{K,0}}&=\sqrt{\lambda},\\
\frac{\rho_{\max}}{\ell}
&=\sqrt{\frac{\lambda}{\mu}}\,
\operatorname{arsinh}\sqrt{\frac{\lambda}{\mu}},
&\frac{|\Delta\phi|}{|\Delta\phi_0|}
&=\sqrt{\lambda}.
\end{aligned}
\label{eq:10:3}
\end{equation}
The angular-sweep ratio applies to either crossing or returning
branches at fixed $\beta$ in the corresponding domain. Increasing
$\lambda$ at fixed $\mu$ makes the optical surface less negatively
curved, enlarges the curvature radius and the Euclidean-embeddable
half-width, and increases the angular sweep. These are comparisons
using the physical length coordinate $\rho$; the rescaling
$x=q\rho$ normalizes the hyperbolic-function argument but transfers
the scale dependence into the radial line element and leaves the
curvature invariant unchanged.

The dependence is summarized in Table~\ref{tab:deformation-trends}. The entries are exact consequences of the metric, rather than fits or perturbative estimates; the sign trends in this table apply to the fixed-$\mu$, fixed-$\ell$ comparison, as specified in the caption. They do not describe a sweep at fixed $\Lambda_e$.
\begin{table*}[t]
\centering
\caption{Exact deformation trends for the fixed-$\mu$, fixed-$\ell$ family at fixed $\beta=|b|/\ell$, referenced to $\lambda=1$. The cosmological parameter varies according to $\Lambda_e=-1/(\lambda\ell^2)$.}
\label{tab:deformation-trends}
\small
\setlength{\tabcolsep}{4pt}
\begin{ruledtabular}
\begin{tabular}{@{}p{0.23\textwidth}p{0.31\textwidth}p{0.19\textwidth}p{0.19\textwidth}@{}}
Observable & Exact dependence & $0<\lambda<1$ & $\lambda>1$\\
\hline
Throat circumference & Eq.~(\ref{eq:4:1}) & unchanged & unchanged\\
Critical impact parameter and ray class & Eqs.~(\ref{eq:5:3})--(\ref{eq:5:4}) & unchanged & unchanged\\
Optical curvature magnitude & Eq.~(\ref{eq:3:1}); ratio $1/\lambda$ & increases & decreases\\
Optical curvature radius & Eq.~(\ref{eq:3:2}); ratio $\sqrt{\lambda}$ & decreases & increases\\
Circumference at fixed $\rho$ & Eq.~(\ref{eq:2:6}) with $q\propto\lambda^{-1/2}$ & decreases & increases\\
Embedding half-width & Eq.~(\ref{eq:6:4}) & decreases & increases\\
Crossing/return angular sweep & Eqs.~(\ref{eq:5:6})--(\ref{eq:5:7}); ratio $\sqrt{\lambda}$ & decreases & increases\\
\end{tabular}
\end{ruledtabular}
\end{table*}

\subsection{Admissible domain and limiting behavior}
\label{subsec:sign-domains}

The metric requires $\lambda>0$, or equivalently
$\ell_{\mathrm{LV}}>-1$. Relative to the undeformed point, the
admissible parameter line separates into
$-1<\ell_{\mathrm{LV}}<0$ ($0<\lambda<1$) and
$\ell_{\mathrm{LV}}>0$ ($\lambda>1$). The boundary
$\ell_{\mathrm{LV}}=-1$ is excluded because $\lambda=0$
degenerates the radial metric coefficient; values below it do not
belong to the static optical family considered here. These are
conditions for this metric ansatz, not universal bounds on all
bumblebee models.

The fixed-parameter trends derived above hold for every finite
$\lambda>0$. The embedding half-width is strictly increasing:
with $y=\sqrt{\lambda/\mu}$,
$\rho_{\max}/\ell=y\operatorname{arsinh}y$ and
\[
\frac{d}{dy}\left(y\operatorname{arsinh}y\right)
=\operatorname{arsinh}y+\frac{y}{\sqrt{1+y^2}}>0.
\]
At the excluded lower endpoint,
$\rho_{\max}/\ell\sim\lambda/\mu$ and $L_K/\ell\to0$
as $\lambda\to0^+$. At the upper endpoint,
$\rho_{\max}/\ell\sim
\sqrt{\lambda/\mu}\ln(2\sqrt{\lambda/\mu})$
and $L_K/\ell\to\infty$ as $\lambda\to\infty$.
These are one-sided limits of the family; neither endpoint is an
admissible finite-parameter geometry.

A complementary protocol holds \(\mu\) and \(\Lambda_e\) fixed while
varying \(\lambda\). Then \(\ell=(-\lambda\Lambda_e)^{-1/2}\) and
\(r_h=\ell\sqrt{\mu}\) vary, whereas
\[
q=\sqrt{-\mu\Lambda_e},\qquad \mathcal K_{\rm opt}=\mu\Lambda_e
\]
remain fixed. Thus intrinsic curvature is unchanged, but the throat
radius and circumference scale as \(\lambda^{-1/2}\), and
\(\chi=q\ell=\sqrt{\mu/\lambda}\) changes. The dimensional crossing
threshold \(b_{\rm crit}=\ell\) and angular sweeps at fixed dimensional
\(b\) therefore do not follow the fixed-\(\ell\), fixed-\(\beta\) trends
of Table~\ref{tab:deformation-trends}. Fixed-\(\ell\) comparisons isolate curvature-scale
changes by allowing \(\Lambda_e\) to vary; fixed-\(\Lambda_e\) comparisons
preserve optical curvature while changing the throat scale. No trend is
protocol-independent unless expressed through the relevant combinations
\(q,\ell,\chi\).

The complementary fixed-$\Lambda_e$ family preserves optical curvature
but changes the throat scale. Its Lorentzian consequences are derived
independently in Sec.~\ref{sec:beyond-optical}.

\section{Beyond optical degeneracy: Lorentzian invariants and probe dynamics}
\label{sec:beyond-optical}

The optical effective-mass degeneracy identified above is exact, but its
scope must be stated precisely. It is a degeneracy of the optical spatial
metric and therefore of the unparameterized spatial paths of null rays.
It does not imply that the original Lorentzian metric, its local curvature
scale, or the dynamics of massive probes are identical for different
values of the Lorentz-violating parameter. Conversely, a quantity that
depends on \(\lambda\) in one parameter protocol need not provide an
independent discriminator in another protocol. We now make these
distinctions explicit using invariant curvature data and probe equations
derived directly from the original metric.

\subsection{Local Lorentzian curvature invariants}

For the metric
\begin{equation}
 ds^2=-f(r)\,dt^2+\frac{\lambda}{f(r)}\,dr^2+r^2d\phi^2,
 \qquad f(r)=\frac{r^2}{\ell^2}-\mu,
 \label{eq:beyond-metric}
\end{equation}
the nonzero diagonal Ricci components reduce to
\begin{equation}
\begin{aligned}
 R_{tt}&=-\frac{2}{\lambda\ell^2}g_{tt},\qquad
 R_{rr}=-\frac{2}{\lambda\ell^2}g_{rr},\\
 R_{\phi\phi}&=-\frac{2}{\lambda\ell^2}g_{\phi\phi}.
\end{aligned}
 \label{eq:beyond-ricci}
\end{equation}
Thus the spacetime is locally of constant negative curvature, with
\begin{equation}
 R_{\mu\nu}=-\frac{2}{\lambda\ell^2}g_{\mu\nu},\qquad
 R=-\frac{6}{\lambda\ell^2},\qquad
 R_{\mu\nu}R^{\mu\nu}=\frac{12}{\lambda^2\ell^4}.
 \label{eq:beyond-invariants}
\end{equation}
In three dimensions the Riemann tensor is algebraically determined by
the Ricci tensor, and the Kretschmann scalar is therefore
\begin{equation}
 R_{\alpha\beta\gamma\delta}R^{\alpha\beta\gamma\delta}
 =\frac{12}{\lambda^2\ell^4}.
 \label{eq:beyond-kretsch}
\end{equation}
These expressions are finite and constant throughout the exterior;
they do not introduce a new curvature singularity at the horizon or
at the reflection throat. The Lorentzian curvature radius is
\begin{equation}
 L_{\mathrm{AdS}}=\sqrt{\lambda}\,\ell,
 \qquad
 R=-\frac{6}{L_{\mathrm{AdS}}^2}.
 \label{eq:beyond-Lads}
\end{equation}
This local invariant scale is distinct from the optical curvature radius
\(\mathcal L_{\mathrm{opt}}=q^{-1}\), because
\begin{equation}
 q^2=\frac{\mu}{\lambda\ell^2},\qquad
 L_{\mathrm{AdS}}^2=\lambda\ell^2,\qquad
 qL_{\mathrm{AdS}}=\sqrt{\mu}.
 \label{eq:beyond-scale-relations}
\end{equation}
The optical curvature therefore depends on the mass-to-deformation
combination \(\mu/\lambda\) at fixed \(\ell\), whereas the Lorentzian
curvature probes the combination \(\lambda\ell^2\).

The local geometry admits a useful exact coordinate interpretation.
With
\begin{equation}
 \mathsf R=\sqrt{\lambda}\,r,\qquad
 \varphi=\frac{\phi}{\sqrt{\lambda}},\qquad
 L_{\mathrm{AdS}}=\sqrt{\lambda}\,\ell,
 \label{eq:beyond-local-map}
\end{equation}
the Lorentzian line element becomes
\begin{equation}
 ds^2=-\left(\frac{\mathsf R^2}{L_{\mathrm{AdS}}^2}-\mu\right)dt^2
 +\frac{d\mathsf R^2}{\mathsf R^2/L_{\mathrm{AdS}}^2-\mu}
 +\mathsf R^2d\varphi^2.
 \label{eq:beyond-canonical-BTZ}
\end{equation}
Locally, this is the standard static BTZ form with AdS radius
\(L_{\mathrm{AdS}}\) and mass parameter \(\mu\). Globally, however,
the original identification \(\phi\sim\phi+2\pi\) becomes
\(\varphi\sim\varphi+2\pi/\sqrt{\lambda}\). The Lorentz-violating
parameter therefore changes the global angular quotient when the
canonical local BTZ coordinates are used. This coordinate result
sharpens the distinction between local curvature and global
identification: at fixed \(\Lambda_e\), local curvature is unchanged,
but the angular period in the canonical chart and the associated
global circumference data still depend on \(\lambda\). At fixed
\(\ell\), both the local AdS curvature scale and the global quotient
change. The optical effective-mass mapping \(M_{\mathrm{opt}}=\mu/\lambda\)
is a separate relation and should not be conflated with this
Lorentzian coordinate equivalence.

The parameter protocol is decisive. At fixed \(\mu\) and \(\ell\),
varying \(\lambda\) changes \(L_{\mathrm{AdS}}\), \(R\), and the
Kretschmann scalar, so local Lorentzian curvature measurements would
distinguish members of the family if \(\ell\) were independently
calibrated. At fixed \(\mu\) and \(\Lambda_e\), however,
\(\ell^{-2}=-\lambda\Lambda_e\) implies
\(L_{\mathrm{AdS}}^2=-1/\Lambda_e\); the local curvature invariants
are then independent of \(\lambda\). In that protocol, curvature
invariants alone cannot identify the deformation. This is not a
contradiction: the two sweeps vary different underlying background
quantities. The statement that Lorentz violation is “optically
degenerate” is exact, while the statement that it is “physically
distinguishable” requires an observable and a specified parameter
calibration.

\subsection{Timelike geodesics: radial dynamics beyond null optics}

Consider a massive test particle moving in the original Lorentzian
exterior, with proper-time parameter \(\tau\). Stationarity and axial
symmetry give the conserved specific energy and angular momentum
\begin{equation}
 E=f(r)\frac{dt}{d\tau},\qquad
 L=r^2\frac{d\phi}{d\tau}.
 \label{eq:beyond-EL}
\end{equation}
The normalization \(u^\mu u_\mu=-1\) yields
\begin{equation}
 \lambda\left(\frac{dr}{d\tau}\right)^2
 =E^2-f(r)\left(1+\frac{L^2}{r^2}\right).
 \label{eq:beyond-timelike}
\end{equation}
The turning-point condition and the circular-orbit equation are
independent of \(\lambda\) when \(E,L,\mu,\ell\) are held fixed,
because \(\lambda\) multiplies the radial kinetic term but does not
enter \(f(r)\). The radial proper-time rate, however, is rescaled by
\(\lambda^{-1/2}\). In coordinate time,
\begin{equation}
 \left(\frac{dr}{dt}\right)^2
 =\frac{f(r)^2}{\lambda E^2}
 \left[E^2-f(r)\left(1+\frac{L^2}{r^2}\right)\right].
 \label{eq:beyond-coordinate-time}
\end{equation}
Thus radial transit times and radial oscillation times can carry
\(\lambda\)-dependence even when the allowed radial interval and
circular-orbit radii do not. Such timing observables are not part of
the optical null-ray data studied in Sec.~\ref{sec:geo}. Their
interpretation requires a specified observer, normalization of
\(t\), and initial data; they should not be confused with an
invariant statement based solely on the optical metric.

\subsection{Massive scalar probe: explicit radial operator}

To probe dynamics beyond null optics, consider a minimally coupled
scalar field of mass \(m_s\) on the Lorentzian metric
Eq.~(\ref{eq:beyond-metric}), with constant \(\lambda>0\). We use the
coordinate time \(t\) of that metric and assume no curvature coupling.
The Klein--Gordon equation is \((\Box-m_s^2)\Psi=0\) \cite{konoplya2011quasinormal},
where
\begin{equation}
  \Box\Psi=
  \frac{1}{\sqrt{-g}}\partial_\mu
  \left(\sqrt{-g}\,g^{\mu\nu}\partial_\nu\Psi\right).
  \label{eq:beyond-dalembertian}
\end{equation}
For Eq.~(\ref{eq:beyond-metric}),
\begin{equation}
  \sqrt{-g}=\sqrt{\lambda}\,r,\qquad
  g^{\mu\nu}=\operatorname{diag}
  \left(-\frac{1}{f},\frac{f}{\lambda},\frac{1}{r^2}\right),
  \label{eq:beyond-KG-metric-data}
\end{equation}
so the field equation becomes
\begin{equation}
  -\frac{1}{f}\partial_t^2\Psi
  +\frac{1}{\lambda r}\partial_r
       \left(rf\,\partial_r\Psi\right)
  +\frac{1}{r^2}\partial_\phi^2\Psi
  -m_s^2\Psi=0.
  \label{eq:beyond-KG-expanded}
\end{equation}
With \(\Psi=e^{-i\omega t+im\phi}R(r)\), \(m\in\mathbb Z\), this reduces exactly to
\begin{equation}
  \frac{1}{r}\frac{d}{dr}
       \left(rf\frac{dR}{dr}\right)
  +\lambda\left[
       \frac{\omega^2}{f}
       -\frac{m^2}{r^2}-m_s^2
       \right]R=0.
  \label{eq:beyond-KG-radial}
\end{equation}
Here \(f=f(r)\), and all radial derivatives are with respect to \(r\). Massive-scalar quasinormal modes have been studied for the static BTZ-like Einstein--bumblebee geometry in Ref.~\cite{ge2025mass}. Equation~(\ref{eq:beyond-KG-radial}) records the corresponding probe
operator for the present metric; we do not calculate its spectrum here. In particular, the appearance of \(\lambda\) alone does not establish a physical spectral shift or break the optical
effective-mass degeneracy. Such a conclusion requires a specified boundary-value problem (including ingoing horizon behavior and an admissible asymptotic condition), together with a fixed normalization of \(t\) for comparisons of \(\omega\). 

\subsection{Degeneracy audit and physical interpretation}

Table~\ref{tab:probe-discriminants} summarizes the logical status of the
principal quantities. “Degenerate” means that the observable is
determined by the same parameter combination as in the effective
static-BTZ optical metric under the stated comparison; it does not
mean that every possible measurement in the full spacetime is
degenerate.

\begin{table*}[t]
\centering
\caption{Probe dependence beyond optical null geometry. Each entry states the comparison protocol and the relevant equation anchor; “candidate discriminator” denotes a derived dependence, not an observational constraint.}
\label{tab:probe-discriminants}
\small
\setlength{\tabcolsep}{4pt}
\begin{ruledtabular}
\begin{tabular}{@{}p{0.18\textwidth}p{0.18\textwidth}p{0.26\textwidth}p{0.26\textwidth}@{}}
Probe or quantity & Parameter protocol & Equation anchor & Physical status\\
\hline
Optical metric and null paths & Fixed $\ell$ & $M_{\rm opt}=\mu/\lambda$, Eq.~(\ref{eq:10:1}) & Exact optical degeneracy\\
Optical curvature & Fixed $\ell$ & Eq.~(\ref{eq:3:1}) & Determined by $M_{\rm opt}$\\
Lorentzian local curvature & Fixed $\ell$ & Eqs.~(\ref{eq:beyond-invariants})--(\ref{eq:beyond-kretsch}) & Separates $\lambda$ only if $\ell$ is calibrated\\
Lorentzian local curvature & Fixed $\Lambda_e$ & Eqs.~(\ref{eq:beyond-invariants})--(\ref{eq:beyond-kretsch}) & Independent of $\lambda$\\
Global angular quotient & Canonical local-BTZ coordinates & Eq.~(\ref{eq:beyond-local-map}) & Global distinction; requires angular-period information\\
Surface gravity & Stated background family & $q$ in Eq.~(\ref{eq:qscale}) & Shares the optical scale; calibration required\\
Timelike radial motion & Fixed $E,L,\mu,\ell$ & Eqs.~(\ref{eq:beyond-timelike})--(\ref{eq:beyond-coordinate-time}) & Candidate timing discriminator\\
Massive scalar probe & Fixed $m_s,\omega,m$ and boundary data & Eq.~(\ref{eq:beyond-KG-radial}) & Candidate spectral discriminator; spectrum not computed\\
Continuum graphene analogue & Fixed optical geometry & Eqs.~(\ref{eq:gr:14})--(\ref{eq:gr:24}) & Inherits optical degeneracy; not a gravitational measurement\\
\end{tabular}
\end{ruledtabular}
\end{table*}

\section{Discussion and conclusions}\label{sec:conc}

The results establish a clear distinction between the Lorentzian
black-hole exterior, its optical geometry, and the two-ended metric
constructed by reflection. The exact optical-coordinate map and ultrastatic
form are fixed by Eqs.~(\ref{eq:2:5})--(\ref{eq:2:10}), and the intrinsic
curvature by Eq.~(\ref{eq:3:1}). The original exterior remains a
one-sided optical half-cylinder: its conformal-boundary end is at
finite optical coordinate, whereas the horizon is at infinite optical
distance. This global structure is not altered by the fact that the
optical profile admits a smooth even continuation.

The reflection construction in Eq.~(\ref{eq:4:1}) produces a smooth,
complete, two-ended ultrastatic geometry with a minimal circle of
radius \(\ell\). Null curves cross this throat in finite affine and
coordinate time, so traversability is established for the prescribed
completed metric. It is not established for the original Lorentzian
black-hole solution. The conditional source calculation in
Eq.~(\ref{eq:4:4}) shows that an interpretation through ordinary
\((2+1)\)-dimensional Einstein gravity with zero cosmological constant
would require a negative radial-null contraction. That diagnostic
does not identify the bumblebee-sector source. A gravitational
wormhole claim would require a separate global solution of the coupled
field equations, including a regular bumblebee configuration and
appropriate boundary conditions.

The null-geodesic first integrals give the exact throat-crossing
threshold, turning points, and complete angular sweeps in
Eqs.~(\ref{eq:5:1})--(\ref{eq:5:8}). The logarithmic divergence at the
critical impact parameter is the separatrix behavior of the optical
centrifugal barrier, not a conventional asymptotic deflection angle.
The finite interval in Eq.~(\ref{eq:6:4}) has a different origin:
it is the reality domain of a Euclidean surface-of-revolution
embedding. It neither truncates the intrinsic optical manifold nor
signals a curvature singularity. The magnetic fields in
Eqs.~(\ref{eq:7:3})--(\ref{eq:7:5}) are external prescriptions on
that embeddable patch. In particular, prescribing a nonzero uniform
signed normal component forces the ambient axial field to diverge at
the throat; this field choice is not equivalent to a regular uniform
ambient axial field.

The continuum Dirac analysis supplies the wave-level result. The
coframe and spin connection determine the covariant operator, whose
radial system reduces to Scarf-II partner equations. The full-line
connection problem yields flux-normalized complex amplitudes and
probabilities, the resonance-pole lattice and residues, spectral
widths and quality factors, and the transmission-phase delay. The
direct numerical comparison of complex amplitudes supports the
analytic connection formulas at the tested real-energy points; it
does not constitute a microscopic-lattice validation or a complete
time-dependent packet simulation. Pole terms in packet dynamics are
conditional on the contour deformation, incident spectral profile,
and observation prescription. They are not a universal exponential
decay law for the full Hermitian continuum problem. A device-level
interpretation additionally requires a lattice deformation, valley
and gauge structure, contacts, and boundary conditions.

The deformation parameter is identifiable only relative to a
specified parameter protocol and observable. At fixed \(\mu\) and
\(\ell\), the optical metric is isometric to the static BTZ optical
metric with effective mass \(M_{\mathrm{opt}}=\mu/\lambda\);
optical-only data therefore constrain this combination rather than
\(\mu\) and \(\lambda\) separately. At fixed \(\mu\) and
\(\Lambda_e\), the optical curvature remains constant while the
throat scale changes. The Lorentzian curvature invariants, timelike
radial dynamics, and massive Klein--Gordon operator derived in
Sec.~\ref{sec:beyond-optical} identify possible probes beyond null
optics, but no probe spectrum or observational bound is computed
here. The appearance of \(\lambda\) in an operator is not itself a
measurement: a discriminator requires calibrated parameters, a fixed
time normalization, and a fully specified initial- or
boundary-value problem.

The contribution is not a new BTZ solution, a general wormhole-support
mechanism, or a new Scarf-II potential. Rather, it is an exact,
parameter-controlled analysis connecting the specified
Lorentz-violating background to its optical geometry, the explicitly
delimited reflection completion, null-ray observables, and continuum
Dirac scattering on that completion. Relative to
earlier BTZ optical and graphene-wormhole analyses, the treatment
retains the deformation factor through the optical-coordinate map
and develops the two-ended Dirac problem through complex amplitudes
and pole-based observables, while making the geometric and material
validity domains explicit. Table~\ref{tab:core-results} collects the
principal results and their domains; Tables~\ref{tab:deformation-trends}
and~\ref{tab:probe-discriminants} distinguish parameter trends from
candidate probes.

The results established here define several concrete directions for further study. A first priority is to construct a sourced Einstein--bumblebee completion as a global boundary-value problem, thereby determining whether the prescribed reflection geometry can arise from a regular solution of the underlying field equations. A second is to calculate Lorentzian probe spectra under explicit AdS boundary conditions and a fixed time normalization, so that the candidate discriminants identified in Sec.~\ref{sec:beyond-optical} can be assessed quantitatively. The continuum scattering predictions should also be tested against a microscopic tight-binding model with specified lattice deformation, contact conditions, and valley structure. Together with the optical-analogue studies of Refs.~\cite{guvendi-1,ahmed-1}, these developments would connect the present exact geometric and continuum analysis to gravitational dynamics, microscopic validation, and experimentally testable observables, while keeping those extensions distinct from the assumptions underlying the results obtained here.


\begin{thebibliography}{99}
\bibitem{BTZ}
M. Ba\~nados, C. Teitelboim, and J. Zanelli,
``The black hole in three-dimensional spacetime,''
Phys. Rev. Lett. \textbf{69}, 1849--1851 (1992),
\href{https://doi.org/10.1103/PhysRevLett.69.1849}
{doi:10.1103/PhysRevLett.69.1849}.

\bibitem{BTZ2}
M. Ba\~nados, M. Henneaux, C. Teitelboim, and J. Zanelli,
``Geometry of the (2+1) black hole,''
Phys. Rev. D \textbf{48}, 1506--1525 (1993),
\href{https://doi.org/10.1103/PhysRevD.48.1506}
{doi:10.1103/PhysRevD.48.1506}.
Erratum: Phys. Rev. D \textbf{88}, 069902 (2013),
\href{https://doi.org/10.1103/PhysRevD.88.069902}
{doi:10.1103/PhysRevD.88.069902}.

\bibitem{BTZEntropy}
S. Carlip,
``The statistical mechanics of the (2+1)-dimensional black hole,''
Phys. Rev. D \textbf{51}, 632--637 (1995),
\href{https://doi.org/10.1103/PhysRevD.51.632}
{doi:10.1103/PhysRevD.51.632}.

\bibitem{Birmingham}
D. Birmingham, I. Sachs, and S. Sen,
``Three-dimensional black holes and string theory,''
Phys. Lett. B \textbf{413}, 281--286 (1997),
\href{https://doi.org/10.1016/S0370-2693(97)01125-8}
{doi:10.1016/S0370-2693(97)01125-8}.

\bibitem{BrownHenneaux}
J. D. Brown and M. Henneaux,
``Central charges in the canonical realization of asymptotic symmetries:
An example from three-dimensional gravity,''
Commun. Math. Phys. \textbf{104}, 207--226 (1986),
\href{https://doi.org/10.1007/BF01211590}
{doi:10.1007/BF01211590}.

\bibitem{CarlipReview}
S. Carlip,
``The (2+1)-dimensional black hole,''
Class. Quantum Grav. \textbf{12}, 2853--2880 (1995),
\href{https://doi.org/10.1088/0264-9381/12/12/005}
{doi:10.1088/0264-9381/12/12/005}.

\bibitem{Mann}
R. B. Mann,
``Black holes of negative mass,''
Class. Quantum Grav. \textbf{14}, 2927--2930 (1997),
\href{https://doi.org/10.1088/0264-9381/14/10/018}
{doi:10.1088/0264-9381/14/10/018}.

\bibitem{Wald}
R. M. Wald, \textit{General Relativity}
(University of Chicago Press, Chicago, 1984).

\bibitem{GibbonsWarnick}
G. W. Gibbons and C. M. Warnick,
``Universal properties of the near-horizon optical geometry,''
Phys. Rev. D \textbf{79}, 064031 (2009),
\href{https://doi.org/10.1103/PhysRevD.79.064031}
{doi:10.1103/PhysRevD.79.064031}.

\bibitem{GibbonsHerdeiro}
G. W. Gibbons, C. A. R. Herdeiro, C. M. Warnick, and M. C. Werner,
``Stationary metrics and optical Zermelo--Randers--Finsler geometry,''
Phys. Rev. D \textbf{79}, 044022 (2009),
\href{https://doi.org/10.1103/PhysRevD.79.044022}
{doi:10.1103/PhysRevD.79.044022}.

\bibitem{CveticGibbons}
M. Cveti\v{c} and G. W. Gibbons,
``Graphene and the Zermelo optical metric of the BTZ black hole,''
Ann. Phys. \textbf{327}, 2617--2626 (2012),
\href{https://doi.org/10.1016/j.aop.2012.05.013}
{doi:10.1016/j.aop.2012.05.013};
\href{https://arxiv.org/abs/1202.2938}{arXiv:1202.2938}.

\bibitem{BTZWormhole}
S. Gurtas Do\u{g}an, A. Guvendi, and O. Mustafa,
``Geometric and wave optics in a BTZ optical metric-based wormhole,''
Phys. Lett. B \textbf{868}, 139824 (2025),
\href{https://doi.org/10.1016/j.physletb.2025.139824}
{doi:10.1016/j.physletb.2025.139824};
\href{https://arxiv.org/abs/2503.18967}{arXiv:2503.18967}.

\bibitem{OpticalWormhole}
S. Gurtas Do\u{g}an, A. Guvendi, and O. Mustafa,
``Ray and wave optics in an optical wormhole,''
Phys. Lett. B \textbf{868}, 139626 (2025),
\href{https://doi.org/10.1016/j.physletb.2025.139626}
{doi:10.1016/j.physletb.2025.139626}.

\bibitem{KosteleckySamuel}
V. A. Kosteleck\'y and S. Samuel,
``Spontaneous breaking of Lorentz symmetry in string theory,''
Phys. Rev. D \textbf{39}, 683--685 (1989),
\href{https://doi.org/10.1103/PhysRevD.39.683}
{doi:10.1103/PhysRevD.39.683}.

\bibitem{KosteleckyPotting}
V. A. Kosteleck\'y and R. Potting,
``CPT and strings,''
Nucl. Phys. B \textbf{359}, 545--570 (1991),
\href{https://doi.org/10.1016/0550-3213(91)90071-5}
{doi:10.1016/0550-3213(91)90071-5}.

\bibitem{Bluhm} R. Bluhm,
``Overview of the Standard Model Extension: Implications and
Phenomenology of Lorentz Violation,''
in \textit{Special Relativity},
edited by J. Ehlers and C. L\"ammerzahl,
Lecture Notes in Physics, Vol. \textbf{702}
(Springer, Berlin, Heidelberg, 2006), pp. 191--226,
\href{https://doi.org/10.1007/3-540-34523-X_8}{doi:10.1007/3-540-34523-X\_8}.

\bibitem{KosteleckyRMP}
V. A. Kosteleck\'y and N. Russell,
``Data tables for Lorentz and CPT violation,''
Rev. Mod. Phys. \textbf{83}, 11--31 (2011),
\href{https://doi.org/10.1103/RevModPhys.83.11}
{doi:10.1103/RevModPhys.83.11}.

\bibitem{BaileyKostelecky}
Q. G. Bailey and V. A. Kosteleck\'y,
``Signals for Lorentz violation in post-Newtonian gravity,''
Phys. Rev. D \textbf{74}, 045001 (2006),
\href{https://doi.org/10.1103/PhysRevD.74.045001}
{doi:10.1103/PhysRevD.74.045001}.

\bibitem{BertolamiParamos}
O. Bertolami and J. P\'aramos,
``Vacuum solutions of a gravity model with vector-induced spontaneous Lorentz symmetry breaking,''
Phys. Rev. D \textbf{72}, 044001 (2005),
\href{https://doi.org/10.1103/PhysRevD.72.044001}
{doi:10.1103/PhysRevD.72.044001}.

\bibitem{Casana}
R. Casana, A. Cavalcante, F. P. Poulis, and E. B. Santos,
``Exact Schwarzschild-like solution in a bumblebee gravity model,''
Phys. Rev. D \textbf{97}, 104001 (2018),
\href{https://doi.org/10.1103/PhysRevD.97.104001}
{doi:10.1103/PhysRevD.97.104001}.

\bibitem{MalufNeves}
R. V. Maluf and J. C. S. Neves,
``Black holes with a cosmological constant in bumblebee gravity,''
Phys. Rev. D \textbf{103}, 044002 (2021),
\href{https://doi.org/10.1103/PhysRevD.103.044002}
{doi:10.1103/PhysRevD.103.044002}.

\bibitem{MaiXuLiangShao}
Z.-F. Mai, R. Xu, D. Liang, and L. Shao,
``Dynamic instability analysis for bumblebee black holes: The odd parity,''
Phys. Rev. D \textbf{109}, 084076 (2024),
\href{https://doi.org/10.1103/PhysRevD.109.084076}
{doi:10.1103/PhysRevD.109.084076}.

\bibitem{MalufAlmeida}
R. V. Maluf, C. A. S. Almeida, R. Casana, and M. M. Ferreira, Jr.,
``Einstein--Hilbert graviton modes modified by the Lorentz-violating bumblebee field,''
Phys. Rev. D \textbf{90}, 025007 (2014),
\href{https://doi.org/10.1103/PhysRevD.90.025007}
{doi:10.1103/PhysRevD.90.025007}.

\bibitem{panah2026optical}
B. Eslam Panah, A. Guvendi, S. Gurtas Dogan, and O. Mustafa,
``Optical and Thermodynamic Signatures of Lorentz Symmetry Breaking in Bumblebee AdS Black Holes,''
Chin. J. Phys. \textbf{103}, 1698-1713 (2026),
\href{https://doi.org/10.1016/j.cjph.2026.07.024}
{doi:10.1016/j.cjph.2026.07.024}.

\bibitem{Ding}
C. Ding, Y. Shi, J. Chen, Y. Zhou, C. Liu, and Y. Xiao,
``Rotating BTZ-like black hole and central charges in
Einstein-bumblebee gravity,''
Eur. Phys. J. C \textbf{83}, 573 (2023),
\href{https://doi.org/10.1140/epjc/s10052-023-11761-y}
{doi:10.1140/epjc/s10052-023-11761-y};
\href{https://arxiv.org/abs/2302.01580}
{arXiv:2302.01580}.

\bibitem{ColladayKostelecky}
D. Colladay and V. A. Kosteleck\'y,
``CPT violation and the standard model,''
Phys. Rev. D \textbf{55}, 6760--6774 (1997),
\href{https://doi.org/10.1103/PhysRevD.55.6760}
{doi:10.1103/PhysRevD.55.6760}.

\bibitem{Kostelecky2004}
V. A. Kosteleck\'y,
``Gravity, Lorentz violation, and the standard model,''
Phys. Rev. D \textbf{69}, 105009 (2004),
\href{https://doi.org/10.1103/PhysRevD.69.105009}
{doi:10.1103/PhysRevD.69.105009}.

\bibitem{KosteleckyTasson}
V. A. Kosteleck\'y and J. D. Tasson,
``Prospects for large relativity violations in matter-gravity couplings,''
Phys. Rev. Lett. \textbf{102}, 010402 (2009),
\href{https://doi.org/10.1103/PhysRevLett.102.010402}
{doi:10.1103/PhysRevLett.102.010402}.

\bibitem{JacobsonMattingly}
T. Jacobson and D. Mattingly,
``Gravity with a dynamical preferred frame,''
Phys. Rev. D \textbf{64}, 024028 (2001),
\href{https://doi.org/10.1103/PhysRevD.64.024028}
{doi:10.1103/PhysRevD.64.024028}.

\bibitem{CarrollFieldJackiw}
S. M. Carroll, G. B. Field, and R. Jackiw,
``Limits on a Lorentz- and parity-violating modification of electrodynamics,''
Phys. Rev. D \textbf{41}, 1231--1240 (1990),
\href{https://doi.org/10.1103/PhysRevD.41.1231}
{doi:10.1103/PhysRevD.41.1231}.

\bibitem{MorrisThorne}
M. S. Morris and K. S. Thorne,
``Wormholes in spacetime and their use for interstellar travel: A tool for teaching general relativity,''
Am. J. Phys. \textbf{56}, 395--412 (1988),
\href{https://doi.org/10.1119/1.15620}{doi:10.1119/1.15620}.

\bibitem{Ellis}
H. G. Ellis,
``Ether flow through a drainhole: A particle model in general relativity,''
J. Math. Phys. \textbf{14}, 104--118 (1973),
\href{https://doi.org/10.1063/1.1666161}
{doi:10.1063/1.1666161}.

\bibitem{Bronnikov}
K. A. Bronnikov,
``Scalar-tensor theory and scalar charge,''
Acta Phys. Pol. B \textbf{4}, 251--266 (1973).

\bibitem{Visser1989}
M. Visser,
``Traversable wormholes: Some simple examples,''
Phys. Rev. D \textbf{39}, 3182(R)--3184(R) (1989),
\href{https://doi.org/10.1103/PhysRevD.39.3182}
{doi:10.1103/PhysRevD.39.3182}.

\bibitem{HochbergVisser1997}
D. Hochberg and M. Visser,
``Geometric structure of the generic static traversable wormhole throat,''
Phys. Rev. D \textbf{56}, 4745--4755 (1997),
\href{https://doi.org/10.1103/PhysRevD.56.4745}
{doi:10.1103/PhysRevD.56.4745}.

\bibitem{HochbergVisser1998}
D. Hochberg and M. Visser,
``The null energy condition in dynamic wormholes,''
Phys. Rev. Lett. \textbf{81}, 746--749 (1998),
\href{https://doi.org/10.1103/PhysRevLett.81.746}
{doi:10.1103/PhysRevLett.81.746}.

\bibitem{FordRoman}
L. H. Ford and T. A. Roman,
``Quantum field theory constrains traversable wormhole geometries,''
Phys. Rev. D \textbf{53}, 5496--5507 (1996),
\href{https://doi.org/10.1103/PhysRevD.53.5496}
{doi:10.1103/PhysRevD.53.5496}.

\bibitem{Lobo}
F. S. N. Lobo,
``Phantom energy traversable wormholes,''
Phys. Rev. D \textbf{71}, 084011 (2005),
\href{https://doi.org/10.1103/PhysRevD.71.084011}
{doi:10.1103/PhysRevD.71.084011}.

\bibitem{VisserBook}
M. Visser,
\textit{Lorentzian Wormholes: From Einstein to Hawking}
(American Institute of Physics Press, Woodbury, NY, 1995).

\bibitem{HawkingEllis}
S. W. Hawking and G. F. R. Ellis,
\textit{The Large Scale Structure of Space-Time}
(Cambridge University Press, Cambridge, 1973).

\bibitem{Perlick}
V. Perlick,
``Gravitational lensing from a spacetime perspective,''
Living Rev. Relativity \textbf{7}, 9 (2004),
\href{https://doi.org/10.12942/lrr-2004-9}
{doi:10.12942/lrr-2004-9}.

\bibitem{GibbonsWerner}
G. W. Gibbons and M. C. Werner,
``Applications of the Gauss--Bonnet theorem to gravitational lensing,''
Class. Quantum Grav. \textbf{25}, 235009 (2008),
\href{https://doi.org/10.1088/0264-9381/25/23/235009}
{doi:10.1088/0264-9381/25/23/235009}.

\bibitem{Werner}
M. C. Werner,
``Gravitational lensing in the Kerr--Randers optical geometry,''
Gen. Relativ. Gravit. \textbf{44}, 3047--3057 (2012),
\href{https://doi.org/10.1007/s10714-012-1458-9}
{doi:10.1007/s10714-012-1458-9}.

\bibitem{Bozza}
V. Bozza,
``Gravitational lensing in the strong field limit,''
Phys. Rev. D \textbf{66}, 103001 (2002),
\href{https://doi.org/10.1103/PhysRevD.66.103001}
{doi:10.1103/PhysRevD.66.103001}.

\bibitem{Ishihara}
A. Ishihara, Y. Suzuki, T. Ono, and H. Asada,
``Finite-distance corrections to the gravitational bending angle
of light in the strong deflection limit,''
Phys. Rev. D \textbf{95}, 044017 (2017),
\href{https://doi.org/10.1103/PhysRevD.95.044017}
{doi:10.1103/PhysRevD.95.044017}.

\bibitem{DoCarmo}
M. P. do Carmo,
\textit{Differential Geometry of Curves and Surfaces:
Revised and Updated Second Edition}
(Dover Publications, Mineola, NY, 2016).

\bibitem{ONeill}
B. O'Neill,
\textit{Semi-Riemannian Geometry with Applications to Relativity}
(Academic Press, New York, 1983).

\bibitem{Jackson}
J. D. Jackson,
\textit{Classical Electrodynamics}, 3rd ed.
(Wiley, New York, 1998).

\bibitem{Carter}
B. Carter,
``Black Hole Equilibrium States,''
in \textit{Black Holes}, edited by C. DeWitt and B. S. DeWitt
(Gordon and Breach, New York, 1973), pp. 57--214.

\bibitem{GrapheneWormhole}
J. Gonz\'alez and J. Herrero,
``Graphene wormholes: A condensed-matter description of Dirac fermions in curved space,''
Nucl. Phys. B \textbf{825}, 426--443 (2010),
\href{https://doi.org/10.1016/j.nuclphysb.2009.09.028}
{doi:10.1016/j.nuclphysb.2009.09.028}.

\bibitem{RojjanasonBurikhamPimsamarn}
T. Rojjanason, P. Burikham, and K. Pimsamarn,
``Charged fermion in (1 + 2)-dimensional wormhole with axial magnetic field,''
Eur. Phys. J. C \textbf{79}, 660 (2019),
\href{https://doi.org/10.1140/epjc/s10052-019-7156-y}
{doi:10.1140/epjc/s10052-019-7156-y}.

\bibitem{NaderiHasanirokh2024}
F. Naderi and K. Hasanirokh,
``Quantum transport of massless Dirac fermions through wormhole-shaped curved graphene in presence of constant axial magnetic flux,''
Sci. Rep. \textbf{14}, 7763 (2024),
\href{https://doi.org/10.1038/s41598-024-57718-3}
{doi:10.1038/s41598-024-57718-3}.

\bibitem{RobertsWiseman2024}
M. M. Roberts and T. Wiseman,
``Analog gravity and continuum effective theory of the graphene tight-binding lattice model,''
Phys. Rev. B \textbf{109}, 045425 (2024),
\href{https://doi.org/10.1103/PhysRevB.109.045425}
{doi:10.1103/PhysRevB.109.045425}.

\bibitem{Gallerati2019}
A. Gallerati,
``Graphene properties from curved space Dirac equation,''
Eur. Phys. J. Plus \textbf{134}, 202 (2019).
doi: \href{https://doi.org/10.1140/epjp/i2019-12610-6}
{10.1140/epjp/i2019-12610-6}.
arXiv:1808.01187.

\bibitem{RobertsWiseman2022}
M. M. Roberts and T. Wiseman,
``Curved-space Dirac description of elastically deformed monolayer graphene is generally incorrect,''
Phys. Rev. B \textbf{105}, 195412 (2022),
\href{https://doi.org/10.1103/PhysRevB.105.195412}
{doi:10.1103/PhysRevB.105.195412}.

\bibitem{Semenoff1984}
G. W. Semenoff,
``Condensed-matter simulation of a three-dimensional anomaly,''
Phys. Rev. Lett. \textbf{53}, 2449--2452 (1984),
\href{https://doi.org/10.1103/PhysRevLett.53.2449}{doi:10.1103/PhysRevLett.53.2449}.

\bibitem{Haldane1988}
F. D. M. Haldane,
``Model for a quantum Hall effect without Landau levels: Condensed-matter realization of the parity anomaly,''
Phys. Rev. Lett. \textbf{61}, 2015--2018 (1988),
\href{https://doi.org/10.1103/PhysRevLett.61.2015}{doi:10.1103/PhysRevLett.61.2015}.

\bibitem{JackiwPi2007}
R. Jackiw and S.-Y. Pi,
``Chiral gauge theory for graphene,''
Phys. Rev. Lett. \textbf{98}, 266402 (2007),
\href{https://doi.org/10.1103/PhysRevLett.98.266402}{doi:10.1103/PhysRevLett.98.266402}.

\bibitem{GuineaKatsnelsonGeim2009}
F. Guinea, M. I. Katsnelson, and A. K. Geim,
``Energy gaps and a zero-field quantum Hall effect in graphene by strain engineering,''
Nat. Phys. \textbf{6}, 30--33 (2010),
\href{https://doi.org/10.1038/nphys1420}{doi:10.1038/nphys1420}.

\bibitem{VozmedianoKatsnelsonGuinea2010}
M. A. H. Vozmediano, M. I. Katsnelson, and F. Guinea,
``Gauge fields in graphene,''
Phys. Rep. \textbf{496}, 109--148 (2010),
\href{https://doi.org/10.1016/j.physrep.2010.07.003}{doi:10.1016/j.physrep.2010.07.003}.

\bibitem{Novoselov2004}
K. S. Novoselov, A. K. Geim, S. V. Morozov, D. Jiang, Y. Zhang,
S. V. Dubonos, I. V. Grigorieva, and A. A. Firsov,
``Electric field effect in atomically thin carbon films,''
Science \textbf{306}, 666--669 (2004),
\href{https://doi.org/10.1126/science.1102896}{doi:10.1126/science.1102896}.

\bibitem{GeimNovoselov2007}
A. K. Geim and K. S. Novoselov,
``The rise of graphene,''
Nat. Mater. \textbf{6}, 183--191 (2007),
\href{https://doi.org/10.1038/nmat1849}{doi:10.1038/nmat1849}.

\bibitem{CastroNeto2009}
A. H. Castro Neto, F. Guinea, N. M. R. Peres, K. S. Novoselov, and A. K. Geim,
``The electronic properties of graphene,''
Rev. Mod. Phys. \textbf{81}, 109--162 (2009),
\href{https://doi.org/10.1103/RevModPhys.81.109}{doi:10.1103/RevModPhys.81.109}.

\bibitem{Thaller1992}
B. Thaller,
\textit{The Dirac Equation}
(Springer, Berlin, 1992).

\bibitem{Novikov2007}
D. S. Novikov,
``Elastic scattering theory and transport in graphene,''
Phys. Rev. B \textbf{76}, 245435 (2007),
\href{https://doi.org/10.1103/PhysRevB.76.245435}{doi:10.1103/PhysRevB.76.245435}.

\bibitem{Katsnelson2006}
M. I. Katsnelson, K. S. Novoselov, and A. K. Geim,
``Chiral tunnelling and the Klein paradox in graphene,''
Nat. Phys. \textbf{2}, 620--625 (2006),
\href{https://doi.org/10.1038/nphys384}{doi:10.1038/nphys384}.

\bibitem{Beenakker2008}
C. W. J. Beenakker,
``Colloquium: Andreev reflection and Klein tunneling in graphene,''
Rev. Mod. Phys. \textbf{80}, 1337--1354 (2008),
\href{https://doi.org/10.1103/RevModPhys.80.1337}{doi:10.1103/RevModPhys.80.1337}.

\bibitem{TaylorScattering}
J. R. Taylor,
\textit{Scattering Theory: The Quantum Theory of Nonrelativistic Collisions}
(Wiley, New York, 1972).

\bibitem{SakuraiNapolitano2017}
J. J. Sakurai and J. Napolitano,
\textit{Modern Quantum Mechanics}, 2nd ed.
(Cambridge University Press, Cambridge, 2017).

\bibitem{LandauLifshitzQM}
L. D. Landau and E. M. Lifshitz,
\textit{Quantum Mechanics: Non-Relativistic Theory}, 3rd ed.
(Pergamon Press, Oxford, 1977).

\bibitem{Morse1929}
P. M. Morse,
``Diatomic molecules according to the wave mechanics. II. Vibrational levels,''
Phys. Rev. \textbf{34}, 57--64 (1929),
\href{https://doi.org/10.1103/PhysRev.34.57}{doi:10.1103/PhysRev.34.57}.

\bibitem{CooperKhareSukhatme1995}
F. Cooper, A. Khare, and U. Sukhatme,
``Supersymmetry and quantum mechanics,''
Phys. Rep. \textbf{251}, 267--385 (1995).
doi: \href{https://doi.org/10.1016/0370-1573(94)00080-M}
{10.1016/0370-1573(94)00080-M}.
arXiv:hep-th/9405029.

\bibitem{KhareSukhatme1988}
A. Khare and U. P. Sukhatme,
``Scattering amplitudes for supersymmetric shape-invariant potentials by operator methods,''
J. Phys. A: Math. Gen. \textbf{21}, L501--L508 (1988),
\href{https://doi.org/10.1088/0305-4470/21/9/005}
{doi:10.1088/0305-4470/21/9/005}.

\bibitem{LevaiEtAl2017}
G. L\'evai, \'A. Baran, P. Salamon, and T. Vertse,
``Analytical solutions for the radial Scarf II potential,''
Phys. Lett. A \textbf{381}, 1936--1942 (2017),
\href{https://doi.org/10.1016/j.physleta.2017.04.010}
{doi:10.1016/j.physleta.2017.04.010}; arXiv:1704.05634.

\bibitem{NISTDLMF}
F. W. J. Olver, A. B. Olde Daalhuis, D. W. Lozier, B. I. Schneider,
R. F. Boisvert, C. W. Clark, B. R. Miller, and B. V. Saunders (eds.),
\textit{NIST Digital Library of Mathematical Functions},
Chapter 15, Hypergeometric Function, especially Eqs.~15.8.2,
15.10.2--15.10.3, and 15.10.6--15.10.7,
National Institute of Standards and Technology,
\href{https://dlmf.nist.gov/15}
{https://dlmf.nist.gov/15}.

\bibitem{Fock1929}
V. Fock,
``Geometrisierung der Diracschen Theorie des Elektrons,''
Z. Phys. \textbf{57}, 261--277 (1929).

\bibitem{Weyl1929}
H. Weyl,
``Elektron und Gravitation,''
Z. Phys. \textbf{56}, 330--352 (1929).

\bibitem{BirrellDavies1982}
N. D. Birrell and P. C. W. Davies,
\textit{Quantum Fields in Curved Space}
(Cambridge University Press, Cambridge, 1982).

\bibitem{ParkerToms2009}
L. Parker and D. Toms,
\textit{Quantum Field Theory in Curved Spacetime: Quantized Fields and Gravity}
(Cambridge University Press, Cambridge, 2009).

\bibitem{VozmedianoDeJuanCortijo2008}
M. A. H. Vozmediano, F. de Juan, and A. Cortijo,
``Gauge fields and curvature in graphene,''
J. Phys.: Conf. Ser. \textbf{129}, 012001 (2008),
\href{https://doi.org/10.1088/1742-6596/129/1/012001}
{doi:10.1088/1742-6596/129/1/012001}.

\bibitem{CortijoVozmediano2007}
A. Cortijo and M. A. H. Vozmediano,
``Effects of topological defects and local curvature on the electronic
properties of planar graphene,''
Nucl. Phys. B \textbf{763}, 293--308 (2007),
\href{https://doi.org/10.1016/j.nuclphysb.2006.10.031}
{doi:10.1016/j.nuclphysb.2006.10.031}.
Erratum: Nucl. Phys. B \textbf{807}, 659--660 (2009),
\href{https://doi.org/10.1016/j.nuclphysb.2008.09.006}
{doi:10.1016/j.nuclphysb.2008.09.006}.

\bibitem{ArminjonReifler2010}
M. Arminjon and F. Reifler,
``Basic quantum mechanics for three Dirac equations in a curved
spacetime,''
Braz. J. Phys. \textbf{40}, 242--255 (2010).
doi: \href{https://doi.org/10.1590/S0103-97332010000200020}
{10.1590/S0103-97332010000200020}.
arXiv:0807.0570 [gr-qc].

\bibitem{OlverASF}
F. W. J. Olver,
\textit{Asymptotics and Special Functions},
Computer Science and Applied Mathematics,
Academic Press, New York, 1974,
Chapter~5, §§8--10, pp.~156--168. \href{https://doi.org/10.1016/B978-0-12-525850-0.50010-9}
{doi:10.1016/B978-0-12-525850-0.50010-9}.

\bibitem{Mostafazadeh2020}
A. Mostafazadeh,
``Transfer matrix in scattering theory: a survey of basic properties and
recent developments,''
Turk. J. Phys. \textbf{44}, 472--527 (2020),
\href{https://doi.org/10.3906/fiz-2009-14}
{doi:10.3906/fiz-2009-14}.

\bibitem{ArovasEtAl2010}
D. P. Arovas, L. Brey, H. A. Fertig, E.-A. Kim, and K. Ziegler,
``Dirac spectrum in piecewise constant one-dimensional (1D) potentials,''
New J. Phys. \textbf{12}, 123020 (2010),
\href{https://doi.org/10.1088/1367-2630/12/12/123020}
{doi:10.1088/1367-2630/12/12/123020}.

\bibitem{TransferMatrixDirac2023}
M. Ibarra-Reyes, R. P\'erez-\'Alvarez, and I. Rodr\'iguez-Vargas,
``Transfer matrix in 1D Dirac-like problems,''
J. Phys.: Condens. Matter \textbf{35}, 395301 (2023),
\href{https://doi.org/10.1088/1361-648X/acdfa2}
{doi:10.1088/1361-648X/acdfa2}.

\bibitem{Newton1982}
R. G. Newton,
\textit{Scattering Theory of Waves and Particles}, 2nd ed.
(Springer-Verlag, New York, 1982),
\href{https://doi.org/10.1007/978-3-642-88128-2}
{doi:10.1007/978-3-642-88128-2}.

\bibitem{CostinSoffer2001}
O. Costin and A. Soffer,
``Resonance theory for Schr\"odinger operators,''
Commun. Math. Phys. \textbf{224}, 133--152 (2001),
\href{https://doi.org/10.1007/s002200100558}
{doi:10.1007/s002200100558}.

\bibitem{Nussenzveig1972}
H. M. Nussenzveig,
``Time delay in quantum scattering,''
Phys. Rev. D \textbf{6}, 1534--1542 (1972),
\href{https://doi.org/10.1103/PhysRevD.6.1534}
{doi:10.1103/PhysRevD.6.1534}.

\bibitem{deCarvalhoNussenzveig2002}
C. A. A. de Carvalho and H. M. Nussenzveig,
``Time delay,''
Phys. Rep. \textbf{364}, 83--174 (2002),
\href{https://doi.org/10.1016/S0370-1573(01)00092-8}
{doi:10.1016/S0370-1573(01)00092-8}.

\bibitem{GarciaCalderonMaldonadoVillavicencio2007}
G. Garc\'ia-Calder\'on, I. Maldonado, and J. Villavicencio,
``Resonant-state expansions and the long-time behavior of quantum decay,''
Phys. Rev. A \textbf{76}, 012103 (2007),
\href{https://doi.org/10.1103/PhysRevA.76.012103}
{doi:10.1103/PhysRevA.76.012103}.

\bibitem{konoplya2011quasinormal}
R. Konoplya, and A. Zhidenko,
``Quasinormal modes of black holes: From astrophysics to string theory,''
Rev. Mod. Phys. \textbf{83}, 793--836 (2011),
\href{https://doi.org/10.1103/RevModPhys.83.793}
{doi:10.1103/RevModPhys.83.793}.


\bibitem{ge2025mass}
F. Ge, Q. Pan, S. Chen, J. Jing,
``Mass ladder operators and quasinormal modes of the static BTZ-like black hole in Einstein-bumblebee gravity,''
Chin. Phys. C \textbf{49}, 015105 (2025),
\href{https://doi.org/10.1088/1674-1137/ad83a9}
{doi:10.1088/1674-1137/ad83a9}.

\bibitem{guvendi-1}
S. Gurtas Dogan, O. Mustafa, A. Guvendi, H. Hassanabadi,
``Optics in spiral dislocation spacetime: Torsion as a geometric waveguide and frequency-filtering mechanism,''
Eur. Phys. J. C \textbf{86}, 31 (2026),
\href{https://doi.org/10.1140/epjc/s10052-025-15239-x}
{doi:10.1140/epjc/s10052-025-15239-x}.

\bibitem{ahmed-1}
F. Ahmed, A. Bouzenada, and {\.I}. Sakall{\i},
``Geometric and wave optics analysis of BTZ-like charged wormholes with disclinations,''
Sci. Rep. (2026),
\href{https://doi.org/10.1038/s41598-026-64171-x}
{doi:10.1038/s41598-026-64171-x}.


\end{thebibliography}
\end{document}